\PassOptionsToPackage{unicode}{hyperref}
\PassOptionsToPackage{hyphens}{url}
\PassOptionsToPackage{dvipsnames,svgnames,x11names}{xcolor}
\documentclass[
  11pt,
  letterpaper,
]{article}
\usepackage{amsmath,amssymb}
\usepackage{iftex}
\ifPDFTeX
  \usepackage[T1]{fontenc}
  \usepackage[utf8]{inputenc}
  \usepackage{textcomp} 
\else 
  \usepackage{unicode-math} 
  \defaultfontfeatures{Scale=MatchLowercase}
  \defaultfontfeatures[\rmfamily]{Ligatures=TeX,Scale=1}
\fi
\usepackage{lmodern}
\ifPDFTeX\else
\fi
\IfFileExists{upquote.sty}{\usepackage{upquote}}{}
\IfFileExists{microtype.sty}{
  \usepackage[]{microtype}
  \UseMicrotypeSet[protrusion]{basicmath} 
}{}
\makeatletter
\@ifundefined{KOMAClassName}{
  \IfFileExists{parskip.sty}{%
    \usepackage{parskip}
  }{
    \setlength{\parindent}{0pt}
    \setlength{\parskip}{6pt plus 2pt minus 1pt}}
}{
  \KOMAoptions{parskip=half}}
\makeatother
\usepackage{xcolor}
\usepackage[margin=1in]{geometry}
\usepackage{color}
\usepackage{fancyvrb}

\DefineVerbatimEnvironment{Highlighting}{Verbatim}{commandchars=\\\{\}}

\usepackage{graphicx}
\usepackage{tikz}
\makeatletter
\def\maxwidth{\ifdim\Gin@nat@width>\linewidth\linewidth\else\Gin@nat@width\fi}
\def\maxheight{\ifdim\Gin@nat@height>\textheight\textheight\else\Gin@nat@height\fi}
\makeatother
\setkeys{Gin}{width=\maxwidth,height=\maxheight,keepaspectratio}
\makeatletter
\def\fps@figure{htbp}
\makeatother
\usepackage{amsmath,amssymb}
\usepackage{microtype}
\usepackage{fvextra}
\usepackage{needspace}
\usepackage{pdflscape}
\usepackage{float}
\usepackage{algorithm}
\usepackage{algpseudocode}
\fvset{fontsize=\small,breaklines=true}
\allowdisplaybreaks[1]
\makeatletter
\renewcommand{\maketitle}{\begin{center}\vspace*{-1em}{\Large\bfseries\@title\par}\vspace{0.8em}{\large\begin{tabular}[t]{c}\@author\end{tabular}\par}\end{center}\vspace{0.4em}}
\makeatother

\usepackage{multirow}
\usepackage{longtable,booktabs,array,calc}
\definecolor{algorithmblue}{HTML}{1F5A85}
\definecolor{hardnessbrown}{HTML}{9B3F19}
\definecolor{thispaperpurple}{HTML}{5B2C83}
\ifLuaTeX
  \usepackage{selnolig}  
\fi
\usepackage{bookmark}
\IfFileExists{xurl.sty}{\usepackage{xurl}}{} 
\hypersetup{
  pdftitle={Small Maximum Likelihood on Fixed Cavender--Farris--Neyman Trees: An Approximation Scheme and Limits to Fully Polynomial Approximation},
  pdfauthor={Mohammad-Hadi Foroughmand-Araabi; Sama Goliaei},
  colorlinks=true,
  linkcolor={blue},
  filecolor={Maroon},
  citecolor={Blue},
  urlcolor={blue},
  pdfcreator={LaTeX via pandoc}}

\title{Small Maximum Likelihood on Fixed Cavender--Farris--Neyman Trees:
An Approximation Scheme and Limits to Fully Polynomial Approximation}
\author{Mohammad-Hadi Foroughmand-Araabi \and Sama Goliaei}
\date{}

\begin{document}
\maketitle

\section{Abstract}\label{abstract}

Small Maximum Likelihood fits the edge substitution probabilities of a
given phylogenetic tree to an observed alignment, summing over the
unobserved ancestral states. For the binary Cavender--Farris--Neyman
model, we assume a uniform root and an edge-specific substitution
probability \(p_e\in[0,\tfrac12]\) for each edge \(e\), shared across
all alignment sites. We give a deterministic polynomial-time
approximation scheme. The algorithm achieves a factor of
\(1+\varepsilon\) for both the negative log likelihood and the loss
after subtracting the uniform-root contribution. Its main step fixes
ancestral sequences at one periodically spaced class of depths while
summing over all remaining ancestors. A posterior entropy bound controls
the cost of this restriction, and dynamic programming optimizes the
resulting components over a finite parameter grid. We also give a
randomized reduction from Clique showing that an FPTAS returning
feasible rational edge probabilities would imply \(RP=NP\). These
results separate fixed-accuracy approximation from fully polynomial
approximation for the same fixed-tree model.

\section{1. Introduction}\label{introduction}

Phylogeny reconstruction seeks an evolutionary tree from sequences
observed at its leaves. A likelihood-based reconstruction must determine
both the tree topology and the parameters describing evolution along its
edges. Even when the topology is supplied, the edge parameters must be
fitted to the data. This problem, called \textbf{Small Maximum
Likelihood} (Small ML), is also the parameter-fitting step used to
evaluate a candidate topology during tree search.

Likelihood evaluation and likelihood optimization are different tasks.
Felsenstein's pruning algorithm efficiently evaluates the likelihood at
specified edge parameters by summing over all ancestral states {[}1{]}.
Finding a parameter vector that globally maximizes this sum remains an
optimization problem. Moreover, each edge parameter is shared across the
alignment, so the sites cannot be optimized independently.
\textbf{Ancestral maximum likelihood} (AML) makes a different choice: it
optimizes a complete assignment of ancestral sequences together with the
edge parameters. Replacing the sum over ancestors by a maximum changes
the objective, and an approximation for AML need not give the same
approximation for Small ML.

We study the binary Cavender--Farris--Neyman (CFN) model on a rooted
binary tree. The root state is uniform, and an edge \(e\) flips its
state with probability \(p_e\in[0,1/2]\). The same vector
\(p=(p_e)_{e\in E}\) applies independently at all \(M\) sites of an
explicit alignment. Let \(L(p)\) be its negative log likelihood and
define the \textbf{normalized loss} \[
F(p)=L(p)-M\ln2.
\] The subtracted term is the contribution of the uniform root. Both
losses are nonnegative and have the same minimizers, but a
multiplicative guarantee for \(F\) is stronger because it controls the
loss beyond this fixed contribution. Write \(L^\star\) and \(F^\star\)
for their minimum values.

\subsection{1.1. Main results}\label{main-results}

\textbf{Theorem 1 (PTAS for Small ML).} Given a rooted binary tree and
an explicit binary alignment, for every rational \(0<\varepsilon\le1\) a
deterministic algorithm returns feasible rational edge probabilities
\(\widehat p\) satisfying \[
F(\widehat p)\le(1+\varepsilon)F^\star,
\qquad
L(\widehat p)\le(1+\varepsilon)L^\star.
\] Its running time is
\(f(1/\varepsilon)N_{\rm in}^{\,2^{O(1/\varepsilon)}}\), where
\(N_{\rm in}\) is the encoded input length and \(f\) depends only on the
accuracy.

Thus Small ML has a polynomial-time approximation scheme (PTAS): the
running time is polynomial in the input length for each fixed accuracy.
A fully polynomial-time approximation scheme (FPTAS) would also be
polynomial in the reciprocal accuracy. Our second result gives a
conditional obstruction to this stronger guarantee.

\textbf{Theorem 2 (Randomized FPTAS obstruction).} If either \(L\) or
\(F\) has an FPTAS that returns feasible rational edge probabilities,
then \(RP=NP\). This holds for explicitly listed alignments on rooted
binary trees, with every edge probability free in \([0,1/2]\).

The reduction in Theorem 2 is randomized; it does not establish
deterministic many-one NP-hardness. Its one-sided error guarantee comes
from extracting and verifying a clique from the returned parameters. The
two theorems are compatible because the exponent in Theorem 1 depends on
\(1/\varepsilon\).

\subsection{1.2. Related work and approximation
landscape}\label{related-work-and-approximation-landscape}

Table 1 separates evaluation at supplied parameters, parameter fitting
on a fixed topology, and optimization over topologies. Here \(n\) is the
number of leaves, \(k\) the number of possible character states at a
site, \(M\) the alignment length, and \(s\) the number of rate
categories. For a symmetric \(k\)-state model with a uniform root, let
\(F_k=L-M\ln k\) be the normalized marginal loss and let
\(J_k=-\sum_i\ln Q_p(z^{(i)})-M\ln k\) be the normalized ancestral loss,
where \(Q_p(z)\) is the largest joint probability of a complete labeling
compatible with \(z\). Thus \(F_2=F\). Write \(J_{\rm u}\) for the
binary ancestral objective allowing \(p_e\in[0,1]\), and
\(J_{\rm full}\) for ancestral loss including the root probability.

In a \textbf{symmetric model}, an edge changes state with probability
\(p_e\), choosing uniformly among the other \(k-1\) states; except in
the explicitly unclipped row, \(0\le p_e\le(k-1)/k\). JC69 is the
four-state case. A \textbf{general time-reversible (GTR) model}
specifies a matrix \(Q\) of instantaneous substitution rates and a
stationary state distribution \(\pi\). Reversibility means
\(\pi_aQ_{ab}=\pi_bQ_{ba}\): at stationarity, the probability flow from
\(a\) to \(b\) equals the reverse flow. The GTR row supplies \(Q\) and
fits branch durations. A \textbf{strict molecular clock} uses a common
evolutionary rate, so each branch length is that rate times the
difference between its endpoint times; the clock row assumes
contemporaneous leaves.

A \textbf{finite rate mixture} assigns each site one unobserved rate
category, shared by all edges at that site; the rates and category
probabilities are supplied. \textbf{Known partitions} instead assign
sites to observed groups, each with its own edge-probability vector and
a common tree. \textbf{No common mechanism} allows a separate
edge-probability vector at every site. \textbf{Parsimony} minimizes the
total state-change cost: \(S(T)\) counts changes with unit costs, and
\(S_c(T)\) uses supplied costs. The near-perfect bound uses
\(S^\star=d+q\), where \(d\) counts nonconstant binary sites.
\textbf{Distance fitting} minimizes squared differences between observed
pairwise distances and tree-path lengths, denoted \(\mathrm{SSE}\).
\textbf{Bayesian inference} targets a posterior distribution after
specifying priors; its integration and sampling guarantees are different
from optimization ratios.

\begin{landscape}
\begingroup
\fontsize{9.2}{10.7}\selectfont
\setlength{\tabcolsep}{4pt}\setlength{\parskip}{0pt}
\renewcommand{\arraystretch}{1.05}
An algorithmic bound \(\rho\le a\) means a polynomial-time \(a\)-approximation for the objective named in the row. A hardness bound \(\rho\ge1+c\) means that some fixed approximation gap exists unless \(P=NP\); \(c>0\) may differ between rows. Subscripts identify the objective when a bound applies to only one of the listed losses. The accuracy \(\varepsilon>0\) is fixed. The GTR and mixture hardness entries concern classes that include the binary symmetric model for the FPTAS obstruction or JC69 for exact topology hardness; the mixture reductions use one category. Appendix C specifies the models and proves the direct consequences used in the table, separating them from the original published statements.

\textbf{Table 1. Algorithmic guarantees and hardness for phylogenetic inference.} Pruning is a bottom-up dynamic program on the tree. Sum-product pruning sums over ancestral states, and max-product pruning finds the highest-probability ancestral assignment. Pruning runtimes count arithmetic operations for the entire alignment. For marginal ML under CFN, we optimize the marginal likelihood over edge probabilities in \([0,\tfrac12]\), with one probability \(p_e\) per edge shared across all alignment sites. Blue \textbf{Algorithm} and brown \textbf{Hardness} labels distinguish upper and lower bounds. \textbf{{[}This paper{]}} marks results proved or derived here; those results use bold purple type. The marking does not assert bibliographic priority. Regular type gives prior guarantees, with ``via'' marking baseline consequences explained in Appendix C. A dash indicates that, to the best of our knowledge, no quantitative approximation guarantee is known for the stated problem and model. ``Not applicable'' identifies an inference task without the stated optimization-ratio interpretation.

\vspace{0.5em}
\begin{tabular}{@{}>{\raggedright\arraybackslash}p{(\linewidth - 8\tabcolsep) * \real{0.095}}>{\raggedright\arraybackslash}p{(\linewidth - 8\tabcolsep) * \real{0.175}}>{\raggedright\arraybackslash}p{(\linewidth - 8\tabcolsep) * \real{0.12}}>{\raggedright\arraybackslash}p{(\linewidth - 8\tabcolsep) * \real{0.305}}>{\raggedright\arraybackslash}p{(\linewidth - 8\tabcolsep) * \real{0.305}}@{}}
\toprule
\textbf{Criterion} & \textbf{Model / objective} & \textbf{Tree and parameters supplied} & \textbf{Optimize on a fixed tree} & \textbf{Optimize the topology} \\ 
\midrule
\multirow[t]{7}{=}{Marginal ML} & Binary CFN; \(F,L\) & \(\Theta(Mn)\) {[}1{]} & \(\textcolor{algorithmblue}{\textsf{Algorithm}}\) \textcolor{thispaperpurple}{\textbf{{[}This paper{]} \(\symbf{\rho\le1+\varepsilon}\)}} (Thm. 1)\newline \(\rho\le2+\varepsilon\) via {[}3{]} (C.2)\newline \(\textcolor{hardnessbrown}{\textsf{Hardness}}\) \textcolor{thispaperpurple}{\textbf{{[}This paper{]} Feasible rational-output FPTAS implies \(RP=NP\)}} (Thm. 2) & \(\textcolor{algorithmblue}{\textsf{Algorithm}}\) \textcolor{thispaperpurple}{\textbf{{[}This paper{]} \(\symbf{\rho\le2\ln4+\varepsilon}\)}} (C.3)\newline \(\rho\le4\) via {[}3{]} (App. B)\newline \(\textcolor{hardnessbrown}{\textsf{Hardness}}\) \(\rho_F\ge1+c\); exact NP-hard {[}4{]} \\[0.7ex]
\cmidrule(l){2-5}
 & JC69 / symmetric fixed \(k\); \(F_k,L\) & \(O(Mnk)\) {[}1{]} & \(\textcolor{algorithmblue}{\textsf{Algorithm}}\) \textcolor{thispaperpurple}{\textbf{{[}This paper{]} \(\symbf{\rho\le1+\varepsilon}\)}} (C.2)\newline \(\rho\le2+\varepsilon\) via {[}3{]} (C.2) & \(\textcolor{algorithmblue}{\textsf{Algorithm}}\) \textcolor{thispaperpurple}{\textbf{{[}This paper{]} \(\symbf{\rho\le2\ln4+\varepsilon}\)}} (C.3)\newline \(\rho\le4\) via {[}3{]} (C.2)\newline \(\textcolor{hardnessbrown}{\textsf{Hardness}}\) Exact NP-hard for JC69 {[}9{]} \\[0.7ex]
\cmidrule(l){2-5}
 & GTR; supplied \(k\)-state generator; \(L\) & \(O(Mnk^2)\) {[}1{]} & \(\textcolor{algorithmblue}{\textsf{Algorithm}}\) \textcolor{thispaperpurple}{\textbf{{[}This paper{]} \(\symbf{\rho\le n+\varepsilon}\)}} (C.4)\newline \(\textcolor{hardnessbrown}{\textsf{Hardness}}\) \textcolor{thispaperpurple}{\textbf{{[}This paper{]} Feasible rational-duration FPTAS implies \(RP=NP\)}} (C.4) & \(\textcolor{algorithmblue}{\textsf{Algorithm}}\) \textcolor{thispaperpurple}{\textbf{{[}This paper{]} \(\symbf{\rho\le n+\varepsilon}\)}} (C.4)\newline \(\textcolor{hardnessbrown}{\textsf{Hardness}}\) Exact NP-hard {[}9{]} \\[0.7ex]
\cmidrule(l){2-5}
 & Strict CFN clock; \(L\) & \(\Theta(Mn)\) {[}1{]} & \(\textcolor{algorithmblue}{\textsf{Algorithm}}\) \textcolor{thispaperpurple}{\textbf{{[}This paper{]} \(\symbf{\rho\le n+\varepsilon}\)}} (C.4) & \(\textcolor{algorithmblue}{\textsf{Algorithm}}\) \textcolor{thispaperpurple}{\textbf{{[}This paper{]} \(\symbf{\rho\le n+\varepsilon}\)}} (C.4)\newline \(\textcolor{hardnessbrown}{\textsf{Hardness}}\) Exact NP-hard {[}10{]} \\[0.7ex]
\cmidrule(l){2-5}
 & Finite positive rate mixture; \(L\) & \(O(Msnk^2)\) {[}1{]} & \(\textcolor{algorithmblue}{\textsf{Algorithm}}\) \textcolor{thispaperpurple}{\textbf{{[}This paper{]} \(\symbf{\rho\le n+\varepsilon}\)}} (C.4)\newline \(\textcolor{hardnessbrown}{\textsf{Hardness}}\) \textcolor{thispaperpurple}{\textbf{{[}This paper{]} Feasible rational-duration FPTAS implies \(RP=NP\)}} (C.4) & \(\textcolor{algorithmblue}{\textsf{Algorithm}}\) \textcolor{thispaperpurple}{\textbf{{[}This paper{]} \(\symbf{\rho\le n+\varepsilon}\)}} (C.4)\newline \(\textcolor{hardnessbrown}{\textsf{Hardness}}\) Exact NP-hard {[}9{]} \\[0.7ex]
\cmidrule(l){2-5}
 & Known CFN partitions; \(F,L\) & \(\Theta(Mn)\) {[}1{]} & \(\textcolor{algorithmblue}{\textsf{Algorithm}}\) \textcolor{thispaperpurple}{\textbf{{[}This paper{]} \(\symbf{\rho\le1+\varepsilon}\)}} (C.5)\newline \(\textcolor{hardnessbrown}{\textsf{Hardness}}\) \textcolor{thispaperpurple}{\textbf{{[}This paper{]} Feasible rational-output FPTAS implies \(RP=NP\)}} (C.5) & \(\textcolor{algorithmblue}{\textsf{Algorithm}}\) \textcolor{thispaperpurple}{\textbf{{[}This paper{]} \(\symbf{\rho\le2\ln4+\varepsilon}\)}} (C.5)\newline \(\textcolor{hardnessbrown}{\textsf{Hardness}}\) \(\rho_F\ge1+c\); exact NP-hard {[}4{]} \\[0.7ex]
\cmidrule(l){2-5}
 & No common mechanism; \(F_k,L\) & \(O(Mnk)\) {[}1{]} & \(\textcolor{algorithmblue}{\textsf{Algorithm}}\) \(\rho=1\) {[}11,12{]} & \(\textcolor{algorithmblue}{\textsf{Algorithm}}\) \(\rho\le\ln4+\varepsilon\) {[}11,18{]} (C.7)\newline \(\textcolor{hardnessbrown}{\textsf{Hardness}}\) \textcolor{thispaperpurple}{\textbf{{[}This paper{]} \(\symbf{\rho\ge1+c}\)}} {[}4,11{]} (C.7) \\[0.7ex]
\bottomrule
\end{tabular}
\endgroup
\end{landscape}

\begin{landscape}
\begingroup
\fontsize{9.2}{10.7}\selectfont
\setlength{\tabcolsep}{4pt}\setlength{\parskip}{0pt}
\renewcommand{\arraystretch}{1.05}
\textbf{Table 1 (continued).} Algorithmic guarantees and hardness for phylogenetic inference. \textcolor{thispaperpurple}{\textbf{[This paper]}} marks results proved or derived here; prior results use regular type.

\vspace{0.5em}
\begin{tabular}{@{}>{\raggedright\arraybackslash}p{(\linewidth - 8\tabcolsep) * \real{0.095}}>{\raggedright\arraybackslash}p{(\linewidth - 8\tabcolsep) * \real{0.175}}>{\raggedright\arraybackslash}p{(\linewidth - 8\tabcolsep) * \real{0.12}}>{\raggedright\arraybackslash}p{(\linewidth - 8\tabcolsep) * \real{0.305}}>{\raggedright\arraybackslash}p{(\linewidth - 8\tabcolsep) * \real{0.305}}@{}}
\toprule
\textbf{Criterion} & \textbf{Model / objective} & \textbf{Tree and parameters supplied} & \textbf{Optimize on a fixed tree} & \textbf{Optimize the topology} \\ 
\midrule
\multirow[t]{4}{=}{Ancestral ML} & Binary CFN; clipped entropy cost; \(J_2\) & \(\Theta(Mn)\) {[}3{]} & \(\textcolor{algorithmblue}{\textsf{Algorithm}}\) \(\rho\le1+\varepsilon\) {[}2,3{]} & \(\textcolor{algorithmblue}{\textsf{Algorithm}}\) \textcolor{thispaperpurple}{\textbf{{[}This paper{]} \(\symbf{\rho\le\ln4+\varepsilon}\)}} (C.3)\newline \(\rho\le2\) {[}3{]}\newline \(\textcolor{hardnessbrown}{\textsf{Hardness}}\) \textcolor{thispaperpurple}{\textbf{{[}This paper{]} \(\symbf{\rho\ge1+c}\)}} (C.7)\newline Exact NP-hard {[}21{]} \\[0.7ex]
\cmidrule(l){2-5}
 & Binary; unclipped entropy cost; \(J_{\rm u}\) & \(\Theta(Mn)\) {[}7{]} & \(\textcolor{algorithmblue}{\textsf{Algorithm}}\) \textcolor{thispaperpurple}{\textbf{{[}This paper{]} \(\symbf{\rho\le1+\varepsilon}\)}} via {[}2{]} (C.2) & \(\textcolor{algorithmblue}{\textsf{Algorithm}}\) \textcolor{thispaperpurple}{\textbf{{[}This paper{]} \(\symbf{\rho\le\ln4+\varepsilon}\)}} (C.3)\newline \(\rho\le16/9\) {[}7{]}\newline \(\textcolor{hardnessbrown}{\textsf{Hardness}}\) \textcolor{thispaperpurple}{\textbf{{[}This paper{]} \(\symbf{\rho\ge1+c}\)}} (C.7)\newline Exact NP-hard {[}21{]} \\[0.7ex]
\cmidrule(l){2-5}
 & JC69 / symmetric fixed \(k\); clipped cost; \(J_k\) & \(O(Mnk)\) {[}3{]} & \(\textcolor{algorithmblue}{\textsf{Algorithm}}\) \(\rho\le1+\varepsilon\) {[}2,3{]} & \(\textcolor{algorithmblue}{\textsf{Algorithm}}\) \textcolor{thispaperpurple}{\textbf{{[}This paper{]} \(\symbf{\rho\le\ln4+\varepsilon}\)}} (C.3)\newline \(\rho\le2\) {[}3{]}\newline \(\textcolor{hardnessbrown}{\textsf{Hardness}}\) \textcolor{thispaperpurple}{\textbf{{[}This paper{]} \(\symbf{\rho\ge1+c}\)}} (C.7)\newline Exact NP-hard via {[}21{]} (C.7) \\[0.7ex]
\cmidrule(l){2-5}
 & Reversible processes / phylogeography; oracle model; \(J_{\rm full}\) & \(O(Mnk^2)\) {[}8{]} & \(\textcolor{algorithmblue}{\textsf{Algorithm}}\) \textcolor{thispaperpurple}{\textbf{{[}This paper{]} \(\symbf{\rho\le n-1}\)}} (C.6) & \(\textcolor{algorithmblue}{\textsf{Algorithm}}\) \textcolor{thispaperpurple}{\textbf{{[}This paper{]} \(\symbf{\rho\le n-1}\)}} (C.6)\newline \(\rho\le2\log_2 n\) {[}8{]}\newline \(\textcolor{hardnessbrown}{\textsf{Hardness}}\) \textcolor{thispaperpurple}{\textbf{{[}This paper{]} \(\symbf{\rho\ge1+c}\)}} (C.7)\newline Exact NP-hard via {[}21{]} (C.7) \\[0.7ex]
\midrule
\multirow[t]{3}{=}{Maximum parsimony} & Unordered, equal costs; \(S\) & \(O(Mnk)\) {[}12{]} & \(\textcolor{algorithmblue}{\textsf{Algorithm}}\) \(\rho=1\) {[}12{]} & \(\textcolor{algorithmblue}{\textsf{Algorithm}}\) \(\rho\le\ln4+\varepsilon\) {[}7,18{]} (C.3)\newline \(\textcolor{hardnessbrown}{\textsf{Hardness}}\) \(\rho\ge1+c\) {[}4{]} \\[0.7ex]
\cmidrule(l){2-5}
 & Supplied metric state-change costs; \(S_c\) & \(O(Mnk^2)\) {[}14{]} & \(\textcolor{algorithmblue}{\textsf{Algorithm}}\) \(\rho=1\) {[}14{]} & \(\textcolor{algorithmblue}{\textsf{Algorithm}}\) \textcolor{thispaperpurple}{\textbf{{[}This paper{]} \(\symbf{\rho\le\ln4+\varepsilon}\)}} (C.3)\newline \(\rho\le2\) {[}18{]}\newline \(\textcolor{hardnessbrown}{\textsf{Hardness}}\) \(\rho\ge1+c\) {[}4{]} \\[0.7ex]
\cmidrule(l){2-5}
 & Binary near-perfect instances; \(S\) & \(O(Mn)\) {[}12{]} & \(\textcolor{algorithmblue}{\textsf{Algorithm}}\) \(\rho=1\) {[}12{]} & \(\textcolor{algorithmblue}{\textsf{Algorithm}}\) \(S_{\rm out}\le d+O(q^2)\)\newline with probability \(\ge1/2\) {[}13{]} \\[0.7ex]
\midrule
Distance fitting & Squared residuals; \(\mathrm{SSE}\) & \(O(n^2)\) & \(\textcolor{algorithmblue}{\textsf{Algorithm}}\) \(\rho=1\) {[}15{]} & \(\textcolor{algorithmblue}{\textsf{Algorithm}}\) \(\rho=O(\log n\log\log n)\) {[}19{]}\newline \(\textcolor{hardnessbrown}{\textsf{Hardness}}\) Exact NP-hard {[}19, §1.1{]} \\[0.7ex]
\midrule
Bayesian inference & Substitution model and priors & \(O(Mnk^2)\) {[}1,17{]} & Not applicable & Not applicable \\[0.7ex]
\bottomrule
\end{tabular}
\endgroup
\end{landscape}

Wang, Jiang, and Lawler {[}2{]} developed a PTAS for metric tree
alignment with a prescribed phylogeny. Elias and Tuller {[}3{]} applied
a related decomposition to Small AML: prescribe observed descendant
sequences at every \(t\)-th depth and optimize the bounded-height
components. We use the same decomposition, but must control its effect
on a sum over ancestral states. Appendix C.2 proves
\(F_k(p)\le J_k(p)\le2F_k(p)\). Consequently, the prior ancestral PTAS
gives a \((2+\varepsilon)\) marginal approximation through this
comparison. Our entropy bound reduces the loss from prescribing a sparse
set of ancestors to \(O(F^\star/t)\), giving the marginal PTAS.

Foroughmand-Araabi, Goliaei, and Alishahi {[}8{]} give a
\(2\log_2 n\)-approximation for ancestral likelihood under supplied
irreducible reversible Markov processes, including phylogeography with
location as an additional evolving coordinate. Their guarantee concerns
the full joint loss and assumes oracles for stationary sequence
probabilities and optimized transition probabilities over a branch
duration. Their node-weighted Steiner-tree formulation uses the
separation of branch optimization once ancestral sequences are chosen.
Small ML sums over ancestral assignments, so that formulation does not
directly apply. Their algorithm also chooses the topology, whereas our
PTAS preserves the supplied tree.

Topology-search hardness and local statistical guarantees address other
tasks. Roch {[}4{]} and Chor--Tuller {[}9,10{]} allow the topology to
vary, so their reductions do not establish fixed-tree hardness. Clancy,
Lyu, and Roch {[}5,6{]} analyze likelihood near the generating
parameters under statistical assumptions. Those results explain when
local optimization succeeds on model-generated data; they do not give a
global approximation guarantee for an arbitrary finite alignment.
Theorem 2 concerns this worst-case fitting problem on a prescribed tree.

\subsection{1.3. Proof overview}\label{proof-overview}

The PTAS first bounds the posterior uncertainty of the ancestral states
by \(F^\star\). Averaged over depth classes and posterior
reconstructions, prescribing the selected sequences costs at most
\(F^\star/t\). A channel-composition argument bounds the average cost of
replacing them by observed descendant sequences by \(6F^\star/t\). Hence
some choice has total additional loss at most \(7F^\star/t\). All other
ancestral states remain summed out. The prescribed vertices divide the
tree into components of height at most \(t\). A finite parameter grid
and dynamic programming find the best solution in this restricted
family.

The hardness proof encodes Clique in a quadratic function of leaf
correlations. We construct a distribution whose expected log likelihood
approximates this function, then sample a polynomial-size alignment that
preserves the gap. The main issue is global control: a Taylor expansion
near the uniform distribution is useful only after ruling out better
fits elsewhere. A relative-entropy argument supplies that control for
every parameter vector whose expected log likelihood is at least that of
the uniform model. Finally, a sufficiently accurate feasible fit yields
a clique that can be checked directly.

\section{2. Model and notation}\label{model-and-notation}

Let \(T=(V,E)\) be a rooted binary tree with root \(\rho\), leaf set
\(\mathcal L\), and internal vertex set \(\mathcal I\). Every internal
vertex has two children. Write \(n=|\mathcal L|\ge2\) and
\(b=|E|=2n-2\). The input alignment has \(M\ge1\) binary sites, also
called loci; a site is one column of the alignment.

We use a subscript for a vertex, edge, or subset and a parenthesized
superscript for a site. Thus \(X_v^{(i)}\) is the random state at vertex
\(v\) and site \(i\), \(X_v=(X_v^{(1)},\ldots,X_v^{(M)})\) is its
sequence, and \(X^{(i)}=(X_v^{(i)})_{v\in V}\) is a full labeling at one
site. Lowercase \(x\) denotes a realization. The observed leaf pattern
at site \(i\) is \(z^{(i)}\), and the observed sequence at leaf \(v\) is
\(z_v\). For \(S\subseteq V\), \(x_S\) restricts a labeling to \(S\); we
use the same convention for \(X\). The shared edge probabilities are
\(p=(p_e)_{e\in E}\), with \(p_{uv}=p_e\) when \(e=uv\). For
\(q\in[0,1/2]\), define \[
K_q(s,t)=
\begin{cases}
1-q,&s=t,\\
q,&s\ne t,
\end{cases}
\qquad s,t\in\{0,1\}.
\] The probability of a complete labeling at one site is \[
\Pr_p\{X^{(i)}=x^{(i)}\}
=\frac12\prod_{uv\in E}K_{p_{uv}}(x_u^{(i)},x_v^{(i)}).
\tag{1}
\]

For a leaf pattern \(z\), let \(P_p(z)\) be the sum of (1) over
compatible internal states. Define the \textbf{uniform-root baseline}
\(B\) for the alignment, the \textbf{negative log likelihood} \(L(p)\),
and the \textbf{normalized loss} \(F(p)\) by \[
B=M\ln2,\qquad
L(p)=-\sum_{i=1}^M\ln P_p(z^{(i)}),\qquad
F(p)=L(p)-B.
\tag{2}
\] All logarithms are natural, with \(-\ln0=+\infty\) and \(0\ln0=0\). A
simultaneous flip of all states preserves probability, so
\(P_p(z)\le1/2\) and \(F(p)\ge0\). The likelihood product is continuous
on the compact box \([0,1/2]^E\) and positive at \(p_e=1/2\) for all
\(e\). Hence a finite optimum is attained.

\section{3. A polynomial-time approximation
scheme}\label{a-polynomial-time-approximation-scheme}

We restrict the ancestral sequences only at selected depths. The proof
first bounds the cost of that restriction, then rounds the edge
probabilities to a finite grid. Dynamic programming optimizes the
restricted problem exactly on the grid.

\subsection{3.1. Posterior entropy}\label{posterior-entropy}

The first lemma bounds ancestral uncertainty by the normalized loss of
one observation. We use entropy \(H_\nu(Y)=-\sum_y\nu(Y=y)\ln\nu(Y=y)\),
with natural logarithms.

\textbf{Lemma 3 (Posterior entropy).} Fix the edge substitution
probabilities \(p=(p_e)_{e\in E}\) and a site \(i\). Observe its leaf
pattern and optionally prescribe states at some internal vertices.
Suppose this observation has probability \(q>0\), and let \(\nu\) be the
conditional distribution of the full labeling. Define \[
f=-\ln(2q),\qquad
\mathcal E=\mathbb E_\nu\sum_{uv\in E}
-\ln K_{p_{uv}}(X_u^{(i)},X_v^{(i)}).
\] Then \[
\sum_{v\in\mathcal I}H_\nu(X_v^{(i)})\le f,
\qquad H_\nu(X_{\mathcal I}^{(i)})\le f,
\qquad \mathcal E\le2f.
\tag{3}
\]

\textbf{Proof.} For a full labeling \(x^{(i)}\) compatible with the
observation, conditioning divides its probability by \(q\). Equation (1)
includes the uniform-root factor \(1/2\), so \[
\nu(x^{(i)})
=\frac{\Pr_p\{X^{(i)}=x^{(i)}\}}{q}
=\frac{\prod_{uv\in E}K_{p_{uv}}(x_u^{(i)},x_v^{(i)})}{2q}.
\] Taking the expected negative logarithm, with the leaf states fixed,
gives \[
f=\mathcal E-H_\nu(X_{\mathcal I}^{(i)}).
\tag{4}
\] Each column of \(K_{p_{uv}}\) is a probability vector. Nonnegativity
of conditional relative entropy therefore implies \[
\mathbb E_\nu[-\ln K_{p_{uv}}(X_u^{(i)},X_v^{(i)})]
\ge H_\nu(X_u^{(i)}\mid X_v^{(i)}).
\tag{5}
\]

The posterior retains the tree factorization: conditional on a parent's
state, its child subtrees are independent. Because the leaves are fixed,
\[
H_\nu(X_{\mathcal I}^{(i)})
=H_\nu(X_\rho^{(i)})+
\sum_{uv\in E}H_\nu(X_v^{(i)}\mid X_u^{(i)}).
\] Apply \(H(X_u\mid X_v)=H(X_v\mid X_u)+H(X_u)-H(X_v)\) to each edge.
Every internal vertex has two children, so the marginal entropy terms
sum to \[
\sum_{uv\in E}H_\nu(X_u^{(i)}\mid X_v^{(i)})
=H_\nu(X_{\mathcal I}^{(i)})+
\sum_{v\in\mathcal I}H_\nu(X_v^{(i)}).
\tag{6}
\] Equations (4)--(6) give the first inequality in (3). Entropy
subadditivity gives the second, and (4) gives the third. All
expectations are over posterior-supported assignments. \(\square\)

\subsection{3.2. Replacing ancestral
sequences}\label{replacing-ancestral-sequences}

For binary sequences \(x,y\) of length \(M\), define the \textbf{clipped
entropy distance} \[
D(x,y)=M H_2\!\left(\min\{d_H(x,y)/M,1/2\}\right),
\qquad H_2(q)=-q\ln q-(1-q)\ln(1-q),
\tag{7}
\] where \(d_H\) is Hamming distance. The function of \(d_H\) in (7),
extended constantly beyond \(M/2\), is nondecreasing, concave, and zero
at zero. Such a function is subadditive, so applying it to Hamming
distance gives a metric. It also equals the minimum transition loss on
an edge whose endpoint sequences are \(x,y\): the fitted probability is
\(\min\{d_H(x,y)/M,1/2\}\).\footnote{With both endpoint sequences fixed,
  let \(d=d_H(x,y)\). The edge likelihood is \(p^d(1-p)^{M-d}\), whose
  maximum on \([0,1]\) occurs at \(p=d/M\). The CFN constraint
  \(p\le1/2\) therefore gives the fitted probability
  \(\min\{d/M,1/2\}\).} For a complete reconstruction \(x\) that agrees
with \(z_v\) at every leaf, define the \textbf{reconstruction cost}
\(R(x)=\sum_{uv\in E}D(x_u,x_v)\). It satisfies \[
R(x)\le\sum_{i=1}^M\sum_{uv\in E}
-\ln K_{p_{uv}}(x_u^{(i)},x_v^{(i)})
\tag{8}
\] for every feasible \(p\).

For selected internal vertices \(S\) with prescribed sequences \(x_S\),
define the \textbf{partial loss} and its normalized version by \[
\begin{aligned}
J_S(p;x_S)
&=-\sum_{i=1}^M\ln\Pr_p\{X_{\mathcal L}^{(i)}=z^{(i)},\ X_S^{(i)}=x_S^{(i)}\},\\
G_S(p;x_S)&=J_S(p;x_S)-B.
\end{aligned}
\tag{9}
\] All unprescribed ancestors are summed out. In particular,
\(L(p)\le J_S(p;x_S)\).

\textbf{Lemma 4 (Sequence replacement).} Suppose no two vertices of
\(S\) are adjacent. For any replacement sequences \(y_S\), some feasible
\(p'\) satisfies \[
J_S(p';y_S)\le J_S(p;x_S)
+\sum_{v\in S}\deg(v)D(x_v,y_v).
\tag{10}
\]

\textbf{Proof.} For \(v\in S\), put \(q_v=\min\{d_H(x_v,y_v)/M,1/2\}\).
On every incident edge, replace \(p_e\) by
\(p_e\oplus q_v=p_e+q_v-2p_eq_v\). These probabilities remain in
\([0,1/2]\), and no edge is modified twice. Binary symmetric channels
satisfy \(K_{p\oplus q}=K_pK_q=K_qK_p\). Keeping one summand in this
product gives, in either edge orientation, \[
K_{p_e\oplus q_v}(y_v^{(i)},s)
\ge K_{q_v}(y_v^{(i)},x_v^{(i)})K_{p_e}(x_v^{(i)},s).
\] Across the alignment, the added negative logarithm is \(D(x_v,y_v)\)
per incident edge. The inequality holds for each assignment of the other
ancestors and therefore survives summation. The root factor remains
\(1/2\). \(\square\)

To choose replacements from the input, let \(d_v(x)\) be the minimum of
\(D(x_v,z_w)\) over descendant leaves \(w\) of \(v\). It is zero at
leaves. For either child \(u\) of \(v\), the triangle inequality gives
\(d_v(x)\le D(x_v,x_u)+d_u(x)\). Summing over the tree yields \[
2\sum_{v\in\mathcal I}d_v(x)
\le R(x)+\sum_{v\in\mathcal I}d_v(x)-d_\rho(x),
\qquad
\sum_{v\in\mathcal I}d_v(x)\le R(x).
\tag{11}
\]

\subsection{3.3. Selecting one depth
class}\label{selecting-one-depth-class}

Fix an integer \(t\ge2\). For \(a\in\{0,\ldots,t-1\}\), let \(S_a\) be
the internal vertices whose depths are congruent to \(a\) modulo \(t\),
with the root at depth zero. Let \(\mathcal S_v\) be the set of observed
sequences at descendant leaves of \(v\).

\textbf{Lemma 5 (Partial reconstruction).} For some residue \(a\), some
sequences \(y_v\in\mathcal S_v\) for \(v\in S_a\), and some feasible
\(p\), \[
G_{S_a}(p;y_{S_a})\le\left(1+\frac7t\right)F^\star.
\tag{12}
\]

\textbf{Proof.} At an optimal vector \(p^\star\), draw the ancestral
states independently across sites from their posteriors given the
observed leaves. Call the resulting reconstruction \(X\). For a fixed
residue, \[
\mathbb E J_{S_a}(p^\star;X_{S_a})
=L^\star+\sum_{i=1}^M H(X_{S_a}^{(i)}\mid z^{(i)}).
\] Entropy subadditivity, averaging over residues, and Lemma 3 give \[
\frac1t\sum_a\mathbb E J_{S_a}(p^\star;X_{S_a})
\le L^\star+\frac{F^\star}{t}.
\tag{13}
\]

At each selected vertex, replace \(X_v\) by a sequence attaining
\(d_v(X)\). Vertices in one depth class are nonadjacent and have degree
at most three. Lemma 4 and (11) bound the average replacement cost by \[
\frac3t\mathbb E\sum_{v\in\mathcal I}d_v(X)
\le\frac3t\mathbb E R(X)
\le\frac6tF^\star,
\] where the last inequality follows from (3) and (8). Adding this bound
to (13) proves (12) for some draw and residue. Sampling proves existence
only; the algorithm searches the observed descendant sequences
deterministically. \(\square\)

\subsection{3.4. A finite parameter grid}\label{a-finite-parameter-grid}

Let \(M_{\rm nc}\) count the nonconstant sites, namely those containing
both leaf states. If \(M_{\rm nc}=0\), setting every \(p_e=0\) attains
\(F^\star=0\). Otherwise a nonconstant pattern specifies different
states at some two leaves \(u,v\), so its probability is at most \[
\frac14\left[1-\prod_{e\in\operatorname{path}(u,v)}(1-2p_e)\right]\le\frac14.
\] Consequently, \[
F^\star\ge M_{\rm nc}\ln2>0.
\tag{14}
\]

We also need to control the total size of the edge probabilities. Every
transition on edge \(e\) has probability at most \(1-p_e\). Apply Lemma
3 to each observation defining the partial loss to obtain \[
M\sum_e p_e
\le M\sum_e[-\ln(1-p_e)]
\le2G_S(p;x_S).
\tag{15}
\] This bound lets us measure the rounding error relative to \(F\),
without paying for the fixed root term.

For rational \(0<\varepsilon\le1\), set \[
\begin{gathered}
t=\left\lceil\frac{14}{\varepsilon}\right\rceil,
\qquad \tau=\frac{\varepsilon}{64},
\qquad \lambda=\frac{\varepsilon M_{\rm nc}}{64Mb},\\
\mathcal G=\{0,1/2\}\cup
\{\lambda(1+\tau)^k:k\ge0,\ \lambda(1+\tau)^k<1/2\}.
\end{gathered}
\tag{16}
\] Round each positive probability upward to the next grid point,
leaving zero unchanged. Its increment obeys
\(\Delta_e\le\tau p_e+\lambda\). Flip probabilities increase, and the
loss of a nonflip transition increases by at most \(2\Delta_e\).
Comparing complete assignments and then summing over unprescribed states
gives \[
\begin{aligned}
G_S(p';x_S)
&\le G_S(p;x_S)+2M\sum_e\Delta_e\\
&\le(1+4\tau)G_S(p;x_S)+2Mb\lambda.
\end{aligned}
\tag{17}
\]

\subsection{3.5. Dynamic programming}\label{dynamic-programming}

Fix a residue \(a\). From each selected vertex, follow descendant edges
until reaching another selected vertex or an original leaf, including
those terminal vertices. If the root is unselected, also include the
component above the first selected vertices. Each edge belongs to one
component, and components meet only at selected vertices. Every
component has height at most \(t\) and fewer than \(H_t=2^{t+1}\)
vertices. Figure 1 illustrates the decomposition.

\begin{figure}[!ht]
\centering
\resizebox{\linewidth}{!}{
\begin{tikzpicture}[
    x=1cm,y=1cm,
    treeedge/.style={line width=1.0pt, line cap=round},
    selected/.style={circle,fill=figblue,draw=figblue,minimum size=5.8pt,inner sep=0pt},
    internal/.style={circle,fill=white,draw=figink,line width=.85pt,minimum size=6.3pt,inner sep=0pt},
    observed/.style={rectangle,fill=white,draw=figink,line width=.85pt,minimum size=6pt,inner sep=0pt},
    captiontxt/.style={font=\scriptsize,text=figgray},
    leaflabel/.style={font=\small,inner sep=1pt}
]
\definecolor{figink}{HTML}{29343C}
\definecolor{figgray}{HTML}{56636E}
\definecolor{figblue}{HTML}{285F91}
\definecolor{figorange}{HTML}{B36A20}
\definecolor{figteal}{HTML}{198374}
\definecolor{figpaleblue}{HTML}{EBF3FA}

\node[anchor=west,font=\bfseries\small] at (0.15,6.65) {(a) Depth classes: $t=2$, $a=0$};
\fill[figpaleblue] (0.42,5.37) rectangle (7.12,5.81);
\fill[figpaleblue] (0.42,2.92) rectangle (7.12,3.36);
\node[captiontxt,anchor=east] at (0.30,6.03) {depth};
\foreach \dy/\txt in {5.59/0,4.37/1,3.14/2,1.93/3}
  \node[captiontxt,anchor=east] at (0.30,\dy) {\txt};

\coordinate (Aroot) at (3.65,5.59);
\coordinate (Au) at (1.88,4.37);
\coordinate (Av) at (5.42,4.37);
\coordinate (Awone) at (1.07,3.14);
\coordinate (Alone) at (2.69,3.14);
\coordinate (Awtwo) at (4.61,3.14);
\coordinate (Altwo) at (6.23,3.14);
\coordinate (Ala) at (.63,1.93);
\coordinate (Alb) at (1.51,1.93);
\coordinate (Alc) at (4.17,1.93);
\coordinate (Ald) at (5.05,1.93);
\foreach \v/\w in {Aroot/Au,Aroot/Av,Au/Awone,Au/Alone,Av/Awtwo,Av/Altwo}
  \draw[treeedge,figblue] (\v)--(\w);
\foreach \v/\w in {Awone/Ala,Awone/Alb}
  \draw[treeedge,figorange] (\v)--(\w);
\foreach \v/\w in {Awtwo/Alc,Awtwo/Ald}
  \draw[treeedge,figteal] (\v)--(\w);
\foreach \v in {Aroot,Awone,Awtwo} \node[selected] at (\v) {};
\foreach \v in {Au,Av} \node[internal] at (\v) {};
\foreach \v in {Alone,Altwo,Ala,Alb,Alc,Ald} \node[observed] at (\v) {};
\node[leaflabel,above right=1pt] at (Aroot) {$\rho$};
\node[leaflabel,above left=0pt] at (Awone) {$w_1$};
\node[leaflabel,above left=0pt] at (Awtwo) {$w_2$};
\node[captiontxt] at (3.65,1.34) {Each edge color marks a different component.};

\node[anchor=west,font=\bfseries\small] at (8.10,6.65) {(b) One component $C$};
\coordinate (Broot) at (12.05,5.59);
\coordinate (Bu) at (10.55,4.37);
\coordinate (Bv) at (13.55,4.37);
\coordinate (Bwone) at (9.82,3.14);
\coordinate (Blone) at (11.28,3.14);
\coordinate (Bwtwo) at (12.82,3.14);
\coordinate (Bltwo) at (14.28,3.14);
\foreach \v/\w in {Broot/Bu,Broot/Bv,Bu/Bwone,Bu/Blone,Bv/Bwtwo,Bv/Bltwo}
  \draw[treeedge,figblue] (\v)--(\w);
\foreach \v in {Broot,Bwone,Bwtwo} \node[selected] at (\v) {};
\foreach \v in {Bu,Bv} \node[internal] at (\v) {};
\foreach \v in {Blone,Bltwo} \node[observed] at (\v) {};
\node[leaflabel,above=7pt] at (Broot) {$v:\,\alpha$};
\node[leaflabel,above left=5pt] at (Bu) {$s\in\{0,1\}$};
\node[leaflabel,above right=5pt] at (Bv) {$s'\in\{0,1\}$};
\node[leaflabel,below=6pt] at (Bwone) {$w_1:\,\beta_{w_1}$};
\node[leaflabel,below=6pt] at (Blone) {$z_{\ell_1}^{(i)}$};
\node[leaflabel,below=6pt] at (Bwtwo) {$w_2:\,\beta_{w_2}$};
\node[leaflabel,below=6pt] at (Bltwo) {$z_{\ell_2}^{(i)}$};
\node[captiontxt] at (12.05,1.88) {Sum over the unprescribed internal states.};
\node[captiontxt] at (12.05,1.48) {Use one edge-parameter vector for all sites.};

\node[selected] at (2.55,.44) {};
\node[captiontxt,anchor=west] at (2.72,.44) {Selected internal vertex};
\node[internal] at (7.15,.44) {};
\node[captiontxt,anchor=west] at (7.32,.44) {Unprescribed internal vertex};
\node[observed] at (12.45,.44) {};
\node[captiontxt,anchor=west] at (12.62,.44) {Observed leaf};
\end{tikzpicture}

}

\textbf{Figure 1.} Left: selected internal vertices at depths congruent
to zero modulo two divide the tree into bounded-height components.
Filled circles are selected vertices, open circles are unprescribed
internal vertices, and squares are original leaves. Right: a component
has upper sequence \(\alpha\) and lower selected sequences
\(\beta_{w_1},\beta_{w_2}\). At site \(i\), pruning fixes the terminal
states and sums over the open-circle states, conditional on
\(\alpha^{(i)}\). One edge-parameter vector is shared across all sites.
\end{figure}

For a component \(C\) with selected upper vertex \(v\), prescribe its
upper sequence \(\alpha\in\mathcal S_v\) and its lower selected
sequences \(\beta_w\in\mathcal S_w\). Let \(\phi_{C,i}\) be the
probability of its terminal symbols at site \(i\), conditional on the
upper symbol. Pruning computes this probability by summing over the
unprescribed internal states. Define \[
c_C(\alpha,(\beta_w))=
\min_{p_{E(C)}\in\mathcal G^{E(C)}}
\left[-\sum_{i=1}^M\ln\phi_{C,i}\right].
\tag{18}
\] The minimization is outside the site sum: every candidate uses one
parameter vector for the whole alignment. For the top component with an
unselected root, sum its two root messages without a factor of \(1/2\).
At a selected upper vertex, use the prescribed state without a prior
factor. The component losses then sum to \(G_{S_a}\), with the
uniform-root contribution removed exactly once.

Let \(C_v\) be the component rooted at a selected vertex \(v\), and
\(B_v\) its lower selected boundary. Define \(V_v(\alpha)\) as the
minimum total loss below \(v\) when its sequence is \(\alpha\). The
recurrence is \[
V_v(\alpha)=
\min_{\beta_w\in\mathcal S_w,\ w\in B_v}
\left[c_{C_v}(\alpha,(\beta_w))+
\sum_{w\in B_v}V_w(\beta_w)\right].
\tag{19}
\] At a selected root, minimize \(V_\rho\) over its candidate sequences.
At an unselected root, combine the top component with the values at its
selected boundary. Store the choices and backtrack to recover the
parameters.

Although (18)--(19) are written as additive losses, their comparisons
require no logarithmic approximation. With rational grid parameters,
pruning returns rational factors. The implementation maximizes their
products across the explicitly listed sites and components, using exact
rational arithmetic. Appendix A bounds the bit lengths. Trying every
residue therefore finds the exact minimum partial loss over the finite
family; Algorithm~\ref{alg:smallml-ptas} summarizes the procedure.

\begin{algorithm}[H]
\caption{\textbf{SmallML-PTAS}}
\label{alg:smallml-ptas}
\begin{algorithmic}[1]
\Require Rooted binary tree $T$, explicit alignment $z$, accuracy $0<\varepsilon\le 1$
\Ensure Feasible edge probabilities $\widehat p$
\If{every alignment column is constant}
    \State \Return $p_e=0$ for every edge $e\in E$
\EndIf
\State Construct $t$ and the parameter grid $\mathcal G$ from (16).
\For{each depth residue $a\in\{0,\ldots,t-1\}$}
    \State Split $T$ into components at the selected vertices $S_a$.
    \For{each component $C$ and each allowed boundary-sequence tuple}
        \State Enumerate the grid vectors $p_{E(C)}\in\mathcal G^{E(C)}$.
        \State Use pruning to compute the likelihood product over all sites.
        \State Store the best product and its parameters.
    \EndFor
    \State Combine the component tables by dynamic programming.
\EndFor
\State Backtrack from the best residue.
\State \Return its edge probabilities $\widehat p$.
\end{algorithmic}
\end{algorithm}

\subsection{3.6. Approximation guarantee and running
time}\label{approximation-guarantee-and-running-time}

\textbf{Proof of Theorem 1.} The case \(F^\star=0\) was handled above.
Otherwise Lemma 5 and the choice of \(t\) give a partial solution with
normalized loss at most \((1+\varepsilon/2)F^\star\). Round its
probabilities using (17). The dynamic program does at least as well as
this grid solution, and removing the prescribed-state constraints only
improves its marginal likelihood. Hence \[
\begin{aligned}
F(\widehat p)
&\le(1+\varepsilon/16)(1+\varepsilon/2)F^\star+2Mb\lambda\\
&\le\left(1+\frac{21\varepsilon}{32}\right)F^\star
\le(1+\varepsilon)F^\star.
\end{aligned}
\tag{20}
\] The second inequality uses \(\varepsilon\le1\), (14), and
\(\ln2\ge1/2\). Adding \(B\) proves the guarantee for \(L\).

The grid has \[
|\mathcal G|=O\!\left(\varepsilon^{-1}
\log\frac{Mb}{\varepsilon M_{\rm nc}}\right)
\tag{21}
\] entries. There are \(O(n)\) components per residue, at most
\(n^{H_t}\) boundary tuples per component, and at most
\(|\mathcal G|^{H_t}\) parameter vectors per tuple. Each evaluation
takes \(O(MH_t)\) pruning operations. The rational arithmetic has
polynomial bit cost for fixed accuracy by Appendix A. Since
\(H_t=2^{O(1/\varepsilon)}\), these bounds give the stated running time
and polynomial-size rational output. \(\square\)

\section{4. Randomized hardness of finer
approximation}\label{randomized-hardness-of-finer-approximation}

We prove Theorem 2 in four steps: encode Clique in a quadratic
correlation score, realize that score through a population likelihood,
sample an explicit alignment, and extract a clique from an accurate
feasible fit. Use the equivalent sign encoding \(\{-1,+1\}\), and write
\(\theta_e=1-2p_e\in[0,1]\). For a leaf pattern
\(X_{\mathcal L}^{(i)}\sim P_p\), the correlation of leaves \(u,v\) is
\[
\rho_{uv}=\mathbb E_p[X_u^{(i)}X_v^{(i)}]
=\prod_{e\in\operatorname{path}(u,v)}\theta_e.
\tag{22}
\]

\subsection{4.1. The graph and tree}\label{the-graph-and-tree}

Start with a Clique instance \((G_0,k_0)\) on \(n_0\) vertices, with
\(1\le k_0\le n_0\). Add \(n_0\) mutually adjacent vertices, each
adjacent to every original vertex. The resulting graph \(G\) has
\(n_G=2n_0\) vertices. Set \(k=k_0+n_0\). Writing \(\omega(G)\) for the
maximum clique size, \[
\omega(G)=\omega(G_0)+n_0,
\qquad \omega(G)\ge n_G/2,
\qquad k\ge n_G/2.
\]

Choose the following parameters, for which \(U,K,N\) and \(1/\eta\) are
polynomial in \(n_G\): \[
U=16n_G^2,\qquad
K=2^{20}Un_G^3,\qquad
N=n_G+U,\qquad
\eta=\frac1{2^{12}N^8K^4}.
\tag{23}
\] The prescribed tree has \(N\) leaves. One root subtree contains the
\(n_G\) \textbf{variable leaves}, one for each graph vertex. The other
contains \(U\) \textbf{anchor leaves}, used to control the correlation
score. Choose the two binary subtree shapes in advance from their leaf
counts. Every edge probability remains free in \([0,1/2]\).

\begin{figure}[H]
\centering
\resizebox{0.84\linewidth}{!}{
\begin{tikzpicture}[x=1cm,y=1cm,
  treeedge/.style={line width=1.3pt,line cap=round},
  labeltext/.style={font=\small},
  subtext/.style={font=\footnotesize,text=figgray}]
\definecolor{figink}{HTML}{29343C}
\definecolor{figgray}{HTML}{56636E}
\definecolor{figblue}{HTML}{285F91}
\definecolor{figteal}{HTML}{198374}
\definecolor{figpaleblue}{HTML}{EBF3FA}
\definecolor{figpaleteal}{HTML}{EAF5F1}
\coordinate (root) at (6.40,5.82);
\coordinate (lefttop) at (3.40,4.20);
\coordinate (righttop) at (9.40,4.20);
\coordinate (leftleaf) at (1.15,1.90);
\coordinate (rightleaf) at (11.65,1.90);

\filldraw[draw=figblue,fill=figpaleblue,line width=1.0pt]
 (lefttop)--(.73,1.87)--(6.07,1.87)--cycle;
\filldraw[draw=figteal,fill=figpaleteal,line width=1.0pt]
 (righttop)--(6.73,1.87)--(12.07,1.87)--cycle;
\draw[treeedge,figblue] (root)--(lefttop);
\draw[treeedge,figteal] (root)--(righttop);
\node[circle,fill=white,draw=figink,line width=1pt,minimum size=7pt,inner sep=0pt] at (root) {};
\node[labeltext,above=9pt] at (root) {Root $\rho$};
\node[labeltext,text=figblue] at (4.43,5.31) {$\gamma_i$};
\node[labeltext,text=figteal] at (8.37,5.31) {$a_j$};
\node[labeltext,text=figblue] at (3.40,2.60) {Variable subtree};
\node[labeltext,text=figteal] at (9.40,2.60) {Anchor subtree};

\draw[figblue,densely dotted,line width=1.0pt] (lefttop)--(leftleaf);
\draw[figteal,densely dotted,line width=1.0pt] (righttop)--(rightleaf);
\node[rectangle,draw=figink,fill=white,minimum size=6pt,inner sep=0pt,line width=1pt] at (leftleaf) {};
\node[rectangle,draw=figink,fill=white,minimum size=6pt,inner sep=0pt,line width=1pt] at (rightleaf) {};
\node[labeltext,above left=2pt] at (leftleaf) {$i$};
\node[labeltext,above right=2pt] at (rightleaf) {$j$};

\node[labeltext] at (3.40,1.30) {Exactly $n_G$ variable leaves};
\node[labeltext] at (9.40,1.30) {Exactly $U$ anchor leaves};
\node[subtext] at (3.40,.83) {One leaf per graph vertex};
\node[subtext] at (9.40,.83) {All anchors lie in this subtree};
\node[font=\normalsize] at (6.40,.17) {$N=n_G+U\qquad \rho_{ij}=\gamma_i a_j$};
\end{tikzpicture}

}

\textbf{Figure 2.} The two root subtrees contain exactly the variable
leaves and the anchor leaves, respectively. The triangles abbreviate
binary subtrees. For variable leaf \(i\) and anchor leaf \(j\), let
\(\gamma_i\) and \(a_j\) be their correlations with the root. Their path
passes through the root, so \(\rho_{ij}=\gamma_i a_j\). All edge
probabilities, including those incident to the root, remain free.
\end{figure}

\subsection{4.2. A quadratic score for
Clique}\label{a-quadratic-score-for-clique}

Scale each pair correlation by \(r_{uv}=\rho_{uv}/\eta\). Let \(A\) be
the symmetric \(N\times N\) matrix indexed by the tree's leaves, with
zero diagonal and \[
A_{uv}=
\begin{cases}
-K,&u,v\text{ are variable leaves forming a nonedge of }G,\\
1,&\text{otherwise}.
\end{cases}
\] Define the \textbf{correlation score} and the \textbf{anchor
baseline} by \[
S(p)=\sum_{\substack{u,v\in\mathcal L\\u<v}}\left(A_{uv}r_{uv}-\frac12r_{uv}^2\right),
\qquad S_0=\frac12\binom U2.
\tag{24}
\] The sum in (24) is over all unordered pairs of distinct leaves of the
tree, with \(|\mathcal L|=N=n_G+U\), including both variable and anchor
leaves. All leaf-pair sums below use this convention. A pair with
coefficient \(1\) contributes at most \(1/2\); a nonedge contributes a
nonpositive penalty.

\begin{samepage}
\textbf{Lemma 6 (Clique score and extraction).} If \(\kappa=\omega(G)\),
then \[
S_0+\frac U2\kappa\le\max_pS(p)
\le S_0+\frac U2\kappa+\frac U8.
\tag{25}
\] From rational parameters satisfying \(S(p)\ge S_0+Un_G/8\), one can
compute in polynomial time a clique \(C\) such that \[
S(p)\le S_0+\frac U2|C|+\frac{5U}{64}.
\tag{26}
\]

\end{samepage}

\textbf{Proof.} Let \(\gamma_i,a_j\) be the root correlations from
Figure 2. Cross-subtree correlations are \(\gamma_i a_j\). Within either
subtree, a leaf-pair correlation is at least the product of its two root
correlations, because that product includes the shared root-path factors
twice. Put \(A_1=\sum_j a_j\) and \(A_2=\sum_j a_j^2\). If \(A_2=0\),
all cross terms vanish and \(S\le S_0+n_G^2/4<S_0+Un_G/8\). Otherwise
define \[
W=\frac{A_1^2}{A_2},\qquad
h_i=\frac{\gamma_iA_2}{\eta A_1}.
\] Then \(1\le W\le U\), and the cross-subtree score is exactly \[
W\sum_i(h_i-h_i^2/2)\le n_GW/2.
\tag{27}
\]

The anchor score is \(S_0-\Delta/2\), where
\(\Delta=\sum_{j<j'}(r_{jj'}-1)^2\). The positive variable-pair terms
contribute at most \(n_G^2/4\). A score below the extraction threshold
already satisfies the upper bound in (25), since \(\kappa\ge n_G/2\). At
or above that threshold, (27) gives \(W\ge U/4-n_G/2\ge U/8\), and
\(S\ge S_0\) gives \[
\Delta+2K\sum_{\rm nonedges}r_{ii'}
\le n_GU+n_G^2/2\le2n_GU.
\tag{28}
\] The within-subtree bound and Cauchy--Schwarz imply \[
A_1^2-A_2
\le2\eta\sum_{j<j'}r_{jj'}
\le\eta[U^2+2U\sqrt{n_GU}]
\le2\eta U^2.
\] As \((W-1)A_2=A_1^2-A_2\) and \(W-1\ge U/16\), we get
\(A_2\le32\eta U\). Hence each variable nonedge satisfies \[
r_{ii'}\ge\frac{\gamma_i\gamma_{i'}}\eta
=\frac{\eta W}{A_2}h_ih_{i'}
\ge\frac1{256}h_ih_{i'}.
\] Combining this with (28) yields \[
\sum_{\rm nonedges}h_ih_{i'}
\le\frac{256n_GU}{K}
=\frac1{4096n_G^2}.
\tag{29}
\]

The set \(C=\{i:h_i\ge1/(16n_G)\}\) is therefore a clique: a nonedge
within \(C\) would contribute at least \(1/(256n_G^2)\) to (29). For
\(i\in C\), use \(h_i-h_i^2/2\le1/2\); otherwise use
\(h_i-h_i^2/2\le h_i<1/(16n_G)\). The cross score is at most
\(U|C|/2+U/16\). Adding the positive variable-pair bound gives \[
S\le S_0+\frac U2|C|+\frac U{16}+\frac{n_G^2}{4}
=S_0+\frac U2|C|+\frac{5U}{64},
\] which proves extraction and the upper bound in (25).

For the lower bound, set all internal edge correlations to one. Give
every anchor pendant edge correlation \(\sqrt\eta\); use the same value
at variable leaves in a maximum clique and zero at all other variable
leaves. The score is \(S_0+U\kappa/2+\kappa(\kappa-1)/4\). These
probabilities are feasible and rational because \(\sqrt\eta\) is
rational. Extraction uses only rational products, sums, divisions, and
comparisons. \(\square\)

\subsection{4.3. From the score to population
likelihood}\label{from-the-score-to-population-likelihood}

The reduction constructs a \textbf{sampling distribution} \(\mu\) from
the graph and draws alignment columns from it. This distribution stays
fixed while likelihood optimization varies the CFN parameters \(p\).

Let \(\mathsf U\) be uniform on \(\{-1,+1\}^N\). Define \[
\begin{aligned}
a(z)&=\sum_{u<v}A_{uv}z_uz_v,\qquad
\mu(z)=2^{-N}(1+\eta a(z)),\\
\mathcal J(p)&=\mathbb E_{Z\sim\mu}\!\left[\ln\!\left(2^NP_p(Z)\right)\right].
\end{aligned}
\tag{30}
\] Thus \(\mathcal J(p)\) averages the log likelihood under \(P_p\) of a
pattern \(Z\sim\mu\), relative to the uniform model. The input alignment
is not assumed to come from a CFN model, so \(\mu\) need not belong to
the family \(\{P_p\}\).

Since \(\mathbb E_{\mathsf U}a=0\), \(|a|\le KN^2\), and
\(\eta KN^2\le1/2\), the function \(\mu\) is a probability distribution
with density between \(1/2\) and \(3/2\) relative to \(\mathsf U\). The
uniform CFN model has \(\mathcal J=0\).

\textbf{Lemma 7 (Global likelihood approximation).} Every parameter
vector with \(\mathcal J(p)\ge0\) satisfies \[
|\mathcal J(p)-\eta^2S(p)|\le\eta^2/32.
\tag{31}
\] The explicit clique parameters in Lemma 6 satisfy the same bound.

\textbf{Proof.} Write
\(D_{\rm KL}(\mu\Vert\nu)=\sum_z\mu(z)\ln(\mu(z)/\nu(z))\) for relative
entropy. The identity \[
\mathcal J(p)=D_{\rm KL}(\mu\Vert\mathsf U)-D_{\rm KL}(\mu\Vert P_p)
\] bounds any model with \(\mathcal J(p)\ge0\). Using \(\ln x\le x-1\)
and orthogonality of the sign monomials, \[
D_{\rm KL}(\mu\Vert\mathsf U)
\le\eta^2\mathbb E_{\mathsf U}a^2
=\eta^2\sum_{u<v}A_{uv}^2.
\] If \(\mathcal J(p)\ge0\), Pinsker's inequality bounds the difference
between any pair moment under \(\mu\) and \(P_p\) by
\(\eta\sqrt{2\sum A_{uv}^2}\). Orthogonality gives
\(\mathbb E_{Z\sim\mu}[Z_uZ_v]=\eta A_{uv}\). In particular, a variable
nonedge has moment \(-\eta K<0\) under \(\mu\), whereas every CFN
correlation in (22) is nonnegative. Therefore \[
0\le\rho_{uv}\le3\eta NK=:c.
\tag{32}
\]

For a leaf set \(A'\), define its moment
\(\rho_{A'}=\mathbb E_p\prod_{v\in A'}X_v^{(i)}\). Odd moments vanish by
global sign symmetry. If \(|A'|=2q\), the moment equals the product of
\(\theta_e\) over edges having an odd number of selected leaves on one
side. In a binary tree, those edges form edge-disjoint paths pairing the
selected leaves. Thus \(\rho_{A'}\) is a product of \(q\) pair
correlations and is at most \(c^q\). The Fourier expansion on sign
patterns is \[
2^NP_p(z)=1+g(z),\qquad
g(z)=\sum_{\varnothing\ne A'\subseteq\mathcal L}
\rho_{A'}\prod_{v\in A'}z_v.
\] The path bound and orthogonality give \[
\|g\|_\infty\le N^2c\le1/4,\qquad
\mathbb E_{\mathsf U}g^2\le N^2c^2,\qquad
\sum_{|A'|\ge4}\rho_{A'}^2\le N^4c^4.
\tag{33}
\] For example, the first sum is bounded by
\(\sum_{q\ge1}(N^2c)^q/(2q)!\le N^2c\); the squared bounds use the same
calculation with \(c^2\).

Now expand \(\ln(1+g)=g-g^2/2+R\), where \(|R|\le|g|^3\) for
\(|g|\le1/4\). The linear term is
\(\mathbb E_{Z\sim\mu}[g(Z)]=\eta\sum A_{uv}\rho_{uv}\). Replacing
\(\mathbb E_{Z\sim\mu}[g(Z)^2]\) by \(\sum\rho_{uv}^2\) costs at most
\(N^4c^4+\eta N^4Kc^2\). By (33) and the density bound on \(\mu\), the
expected remainder is at most \(2N^4c^3\). The total error is bounded by
\[
\frac12N^4c^4+\frac12\eta N^4Kc^2+2N^4c^3
\le3N^4c^3
=81\eta^3N^7K^3
\le\eta^2/32.
\] The remaining quadratic expression is \(\eta^2S(p)\). For the
explicit clique parameters, all pair correlations are at most \(\eta\),
so the same expansion applies directly. \(\square\)

Define the YES and NO thresholds and the gap scale by \[
\begin{aligned}
J_{\rm yes}&=\eta^2\left(S_0+\frac{Uk}{2}-\frac1{32}\right),\\
J_{\rm no}&=\eta^2\left(S_0+\frac{U(k-1)}2+\frac U8+\frac1{32}\right),\\
g_0&=\frac{\eta^2U}{4}.
\end{aligned}
\tag{34}
\] Lemmas 6--7 imply \[
\begin{aligned}
\omega(G)\ge k&\ \Longrightarrow\ \max_p\mathcal J(p)\ge J_{\rm yes},\\
\omega(G)\le k-1&\ \Longrightarrow\ \max_p\mathcal J(p)\le J_{\rm no},\\
J_{\rm yes}-J_{\rm no}&\ge g_0.
\end{aligned}
\tag{35}
\] Candidates with negative \(\mathcal J\) lie below \(J_{\rm no}\). The
reciprocal gap is polynomial in the source size. The distribution has
exponentially many patterns, however, so we still need a polynomial-size
alignment.

\subsection{4.4. From population likelihood to an
alignment}\label{from-population-likelihood-to-an-alignment}

For this tree, let \(b=2N-2\). For patterns \(Z^{(1)},\ldots,Z^{(M)}\)
sampled independently from \(\mu\), define the population and empirical
mean losses \[
\begin{aligned}
\overline L_\mu(p)&=-\mathbb E_{Z\sim\mu}[\ln P_p(Z)]
=N\ln2-\mathcal J(p),\\
\overline L_{\rm emp}(p)&=-\frac1M\sum_{i=1}^M\ln P_p(Z^{(i)}).
\end{aligned}
\] Set the clipping level and net spacing to \[
\delta=\frac{g_0}{128b},\qquad
h=\frac{g_0\delta}{128b}.
\tag{36}
\]

\textbf{Lemma 8 (Explicit alignment).} A randomized procedure with
polynomial worst-case running time either reports failure or outputs a
polynomial-size alignment. With probability greater than \(0.99\), it
outputs an alignment satisfying \[
\sup_{p\in[\delta,1/2]^b}
|\overline L_{\rm emp}(p)-\overline L_\mu(p)|\le g_0/16.
\tag{37}
\] Clipping every coordinate below \(\delta\) to \(\delta\) raises
either mean loss by at most \(g_0/32\).

\textbf{Proof.} Clipping increases flip probabilities and reduces each
nonflip factor by a ratio no smaller than \(1-2\delta\). Its loss
increase is at most \(-b\ln(1-2\delta)\le4b\delta=g_0/32\). On the
clipped box, every pattern has probability at least \(\delta^b/2\), so
its loss is bounded by \(R_0=\ln2+b\ln(1/\delta)\). Each loss derivative
is a posterior average of a log-transition derivative and has absolute
value at most \(1/\delta\).

An infinity-norm net of mesh \(h\) has at most \((2/h)^b\) points.
Moving to a net point changes a pattern loss by at most
\(bh/\delta=g_0/128\). Hoeffding's inequality and a union bound give
error at most \(g_0/32\) on the net, with failure probability at most
\(1/400\), when \[
M\ge\max\left\{16,\frac{512R_0^2}{g_0^2}
\left[b\ln(2/h)+\ln800\right]\right\}.
\tag{38}
\] The two off-net comparisons add \(2g_0/128\), proving (37).
Computable integer upper bounds on the logarithms select a polynomial
sample size. The net is used only in the proof.

To sample exactly, write \(\eta=1/Q\) and \(W_A=\sum_{u<v}|A_{uv}|\).
Draw a uniform sign vector \(z\) and accept it with probability
\((Q+a(z))/(Q+W_A)\). Accepted vectors have law \(\mu\), and each trial
succeeds with probability \(Q/(Q+W_A)\ge2/3\). Allow \(4M\) trials,
reporting failure if fewer than \(M\) are accepted. Chernoff's
inequality bounds that failure probability by \(e^{-25M/48}<1/400\).
Implement each acceptance coin by a uniform integer in
\(\{0,\ldots,Q+W_A-1\}\), using rejection from a power-of-two interval.
Cap each integer draw at \(\lceil\log_2(1600M)\rceil\) attempts and
report failure if its cap is exceeded. Each attempt succeeds with
probability at least \(1/2\), so all \(4M\) draws together fail with
probability at most \(1/400\). Coupling to the unlimited exact sampler,
the concentration event and both runtime caps succeed with probability
at least \(1-3/400>0.99\). Every loop and integer encoding has
polynomial size. \(\square\)

\subsection{4.5. Extracting and verifying a
clique}\label{extracting-and-verifying-a-clique}

\textbf{Proof of Theorem 2.} Suppose an FPTAS for \(L\) is available.
Generate the alignment of Lemma 8 and reject if sampling reports
failure. Let \(\mathrm{OPT}_{\rm emp}\) and \(\mathrm{OPT}_\mu\) be the
unrestricted optimal mean losses. Run the FPTAS with \[
\alpha=\frac{g_0}{64N}.
\tag{39}
\] The uniform model gives \(\mathrm{OPT}_{\rm emp}\le N\ln2\le N\).
Therefore the returned \(\widehat p\) has empirical mean loss at most
\(\mathrm{OPT}_{\rm emp}+g_0/64\). The alignment length and \(1/\alpha\)
are polynomial in the source size.

Let \(\bar p\) be the clipped version of \(\widehat p\). Compare it to a
clipped population optimizer. On (37), two empirical-to-population
comparisons, two clipping bounds, and the FPTAS guarantee yield \[
\overline L_\mu(\bar p)
\le\mathrm{OPT}_\mu+
2\frac{g_0}{16}+2\frac{g_0}{32}+\frac{g_0}{64}
=\mathrm{OPT}_\mu+\frac{13g_0}{64}
<\mathrm{OPT}_\mu+\frac{g_0}{4}.
\tag{40}
\] On a YES instance, \(\mathcal J(\bar p)\ge J_{\rm yes}-g_0/4>0\).
Lemma 7 then gives \[
S(\bar p)\ge S_0+\frac{Uk}{2}-\frac U{16}-\frac1{16}.
\tag{41}
\] This exceeds the extraction threshold in Lemma 6. That lemma produces
a clique \(C\); if \(|C|\le k-1\), its upper bound would contradict
(41). Thus \(|C|\ge k\).

Remove the added universal vertices, leaving at least \(k_0\) original
vertices. Compute the \(h_i\) and \(C\) by exact rational arithmetic,
rejecting if a required denominator vanishes, and accept only after
checking the size and every edge of this original clique. A NO instance
is never accepted, including when the sampling event fails. A YES
instance is accepted with probability greater than \(0.99\). This gives
an RP algorithm for Clique and hence \(RP=NP\).

For an FPTAS for \(F\), the uniform model bounds the optimal normalized
mean loss by \((N-1)\ln2\le N\). The same accuracy \(\alpha\) therefore
gives the same additive error \(g_0/64\). Subtracting the fixed root
term leaves the comparison (40) unchanged, so the argument applies to
\(F\) as well. \(\square\)

The feasible output is what gives one-sided error. For comparison, (37)
and the clipping bound imply
\(|\mathrm{OPT}_{\rm emp}-\mathrm{OPT}_\mu|\le3g_0/32\): the clipped
optima differ by at most \(g_0/16\), and each clipping increase lies in
\([0,g_0/32]\). Thus the empirical YES and NO intervals remain separated
by at least \(13g_0/16\). A polynomial-time value-only algorithm with
mean additive error at most \(g_0/16\) would distinguish them with
bounded two-sided error, implying \(NP\subseteq BPP\). The threshold
involving \(N\ln2\) can be evaluated to error \(g_0/100\) in polynomial
time.

\section{5. Discussion}\label{discussion}

The PTAS gives a worst-case approximation guarantee, but its dependence
on accuracy is too large for direct practical use. The main cost is the
enumeration of boundary sequences and parameters in components of height
\(O(1/\varepsilon)\). Reducing this dependence is a natural next
question. The FPTAS obstruction leaves open an efficient PTAS with
running time \(f(1/\varepsilon)\operatorname{poly}(N_{\rm in})\).

The hardness reduction is deterministic until it converts a distribution
into an alignment. A deterministic construction preserving the global
likelihood gap would strengthen the conclusion. Matching only a few
moments does not supply the uniform loss bound used here. The result
also concerns optimization of the supplied alignment: the reduction's
sampling distribution need not be a CFN distribution, so it does not
contradict parameter-recovery guarantees for CFN-generated data.

The same approximation scheme extends to an input listing distinct site
patterns with positive integer multiplicities encoded in binary.
Interpret \(M\) and \(M_{\rm nc}\) as the total and nonconstant counts.
Because every prescribed boundary sequence is copied from an observed
leaf, it agrees across repeated occurrences of a pattern; pruning can
therefore score each distinct pattern once. Exact likelihood products
may now have excessive bit length, so instead weight certified rational
log approximations by the counts. Per-pattern, per-component error at
most \(\xi=\varepsilon M_{\rm nc}/(128Mb)\) gives global comparison loss
at most \(2Mb\xi\le\varepsilon F^\star/32\), within the slack in (20).
Rational logarithms can be approximated in time polynomial in their bit
length and \(\log(1/\xi)\), so the running time remains polynomial in
the compressed input length for fixed accuracy.

Appendix C.2 extends the PTAS to fixed symmetric alphabets. For more
general evolutionary models, the proof identifies three ingredients to
examine: an entropy bound for the conditional ancestral states, a
sequence-replacement inequality compatible with the transition channels,
and a parameter grid whose loss is controlled after removing the root
contribution. Any such extension must retain the sum over the
unprescribed ancestors.

\section{AI-use disclosure}\label{ai-use-disclosure}

ChatGPT generated the mathematical arguments in this manuscript,
including the proof of the approximation scheme and the randomized
hardness reduction.

\section{Appendix A. Exact
arithmetic}\label{appendix-a.-exact-arithmetic}

The grid in (16) consists of rational powers of \(1+\tau\), scaled by
\(\lambda\). A point \(\lambda(1+\tau)^k\) has bit length bounded by the
bit length of \(\lambda\) plus \(k\) times that of \(1+\tau\). Equation
(21) bounds \(k\), so these lengths are polynomial in the input length
for fixed accuracy. If each grid parameter has at most \(s\) bits, a
component factor is a sum of at most \(2^{H_t}\) products of at most
\(H_t\) transitions. The product of the edge denominators is a common
denominator, giving numerator and denominator lengths \(O(H_t(s+1))\).
Pruning computes these factors exactly and detects zero exactly.

For an explicit alignment, multiplying over its \(M\) sites gives
component products of \(O(MH_t(s+1))\) bits. Multiplying over all
components gives \(O(Mb(s+1))\) bits. Exact comparisons use integer
cross multiplication, and the dynamic program stores the maximizing
choices. Intermediate products are bounded in the same way because the
components have disjoint edge sets. Backtracking returns the selected
rational grid parameters, with no transcendental comparisons.

\section{Appendix B. Marginal and ancestral
likelihood}\label{appendix-b.-marginal-and-ancestral-likelihood}

Let \(Q_p(z)\) be the probability of the most likely complete labeling
compatible with a leaf pattern \(z\). The largest posterior mass is at
least the exponential of minus its entropy. Lemma 3 bounds that entropy
by \(-\ln(2P_p(z))\), yielding \[
2P_p(z)^2\le Q_p(z)\le P_p(z).
\tag{42}
\] Define the normalized ancestral loss
\(J(p)=-\sum_i\ln Q_p(z^{(i)})-B\). Equation (42) gives
\(F(p)\le J(p)\le2F(p)\), and therefore
\(F^\star\le J^\star\le2F^\star\). An ancestral approximation
consequently incurs an additional factor of at most two when transferred
through this comparison. Lemma 5 avoids that constant loss by
prescribing only one depth class.

The same comparison gives the factor-four topology bound retained in
Table 1 for comparison with the stronger consequence in Appendix C.3.
Let \(R_{\rm all}\) be the minimum reconstruction cost over all binary
topologies and full ancestral labelings, and \(F_{\rm all}^\star\) the
minimum marginal loss over topologies. Then
\(R_{\rm all}\le2F_{\rm all}^\star\). Construct a minimum spanning tree
on the observed sequences using the metric \(D\). Doubling an optimal
reconstruction tree and shortcutting between observed leaves bounds its
cost by \(2R_{\rm all}\). This is the standard metric spanning-tree
approach to ancestral reconstruction {[}3, 7{]}.

Attach a named leaf with the same sequence at each observed-sequence
vertex, suppress degree-two internal vertices, and resolve higher
degrees with copies of the same sequence on zero-cost edges. Subdivide
an edge and give the new binary root the sequence of one endpoint. These
operations produce the required named leaves without increasing cost.
Fit each edge by its clipped disagreement fraction. The marginal
likelihood includes that reconstruction, so the returned tree
\(\widehat T\) and parameters satisfy \[
F_{\widehat T}(\widehat p)\le2R_{\rm all}\le4F_{\rm all}^\star.
\] The same factor follows for \(L\) after adding \(B\). Because \(D\)
is nondecreasing in Hamming distance, Kruskal's algorithm can use
integer Hamming distances to order the edges, and the fitted
probabilities are rational. This proves the stated polynomial-time upper
bound.

\section{Appendix C. Assumptions and derivations for the comparison
table}\label{appendix-c.-assumptions-and-derivations-for-the-comparison-table}

\subsection{C.1. Model assumptions and source
scope}\label{c.1.-model-assumptions-and-source-scope}

Table C.1 gives the assumptions behind each row of Table 1. The results
below that table establish the additional bounds used in the comparison.
A consequence proved here is distinguished from a theorem stated in the
cited paper; no claim of bibliographic priority is made for these
consequences.

\begin{landscape}
\begingroup
\fontsize{9.5}{11.2}\selectfont
\setlength{\tabcolsep}{4pt}\setlength{\parskip}{0pt}
\renewcommand{\arraystretch}{1.05}
\textbf{Table C.1. Detailed scope of the comparison.} The rows follow Table 1 in the same order. The likelihood models use independent alignment sites; their shared parameters are specified separately in each row.

\vspace{0.5em}
\begin{tabular}{@{}>{\raggedright\arraybackslash}p{(\linewidth - 4\tabcolsep) * \real{0.18}}>{\raggedright\arraybackslash}p{(\linewidth - 4\tabcolsep) * \real{0.35}}>{\raggedright\arraybackslash}p{(\linewidth - 4\tabcolsep) * \real{0.47}}@{}}
\toprule
\textbf{Problem and model} & \textbf{Optimization variables and objective} & \textbf{Algorithms, sources, and qualifications} \\ 
\midrule
Marginal CFN & Uniform binary root; one probability \(p_e\in[0,1/2]\) per edge, shared across sites. Optimize \(F=L-M\ln2\) or \(L\). & Sum-product pruning evaluates a supplied vector {[}1{]}. Theorems 1--2 concern an explicit alignment and feasible rational parameter output. The prior ancestral PTAS {[}3{]} gives only \(2+\varepsilon\) through C.2. Roch's topology gap applies to \(F\); its transfer to full \(L\) is not asserted {[}4{]}. \\[0.7ex]
\midrule
Marginal symmetric \(k\)-state / JC69 & Fixed \(k\), uniform root, total change probability \(p_e\in[0,(k-1)/k]\), and probability \(p_e/(k-1)\) for each particular change. Optimize \(F_k\) or \(L\). & Symmetry reduces each pruning update to a total over child states. C.2 extends the PTAS; it does not extend Theorem 2's hardness. The exact topology hardness cited in Table 1 is for JC69 {[}9{]}, not a separate claim for every fixed \(k\). \\[0.7ex]
\midrule
Marginal supplied GTR & A rational irreducible reversible generator \(Q\) and its stationary root law are supplied. Fit finite nonnegative branch durations with no upper bound. Optimize full \(L\), relative to its infimum. & C.4 gives finite rational durations. Fitting a supplied generator differs from jointly fitting its rates and stationary law. The FPTAS obstruction is for the model class allowing \(k=2\); JC69 supplies the exact topology hardness {[}9{]}. Neither statement establishes hardness for every fixed generator. \\[0.7ex]
\midrule
Marginal strict CFN clock & A common substitution rate and contemporaneous leaves. Fit internal node times, with nonnegative induced branch durations and no calibration or upper bounds. Optimize full \(L\). & Shared node times couple branch lengths. The construction in C.4 has equal pendant durations and zero internal durations, so it respects the clock. Chor--Tuller {[}10{]} prove exact topology hardness under a molecular clock. No fixed-tree hardness is inferred from that result. \\[0.7ex]
\midrule
Marginal finite rate mixture & Supply \(Q\), positive rational rates \(r_1,\ldots,r_s\), and positive rational weights summing to one. One latent category applies to every edge at a site; all categories use the same duration vector. Optimize full \(L\). & Prune separately for each category and average its likelihood {[}1{]}. The row does not jointly fit rates or weights. C.4 uses the smallest supplied rate. Hardness restricts the model class to \(s=1\), with \(k=2\) for the FPTAS obstruction and JC69 for exact topology hardness {[}9{]}; no claim is made for prescribed distinct categories with \(s>1\). \\[0.7ex]
\midrule
Marginal known CFN partitions & The partition of sites is supplied. Each group has its own CFN edge vector, and all groups use the same topology. Sum their normalized or full losses. & Fixed-tree optimization separates by group. Topology search does not: C.5 combines group costs on a common component topology. Both hardness statements follow by taking one partition. \\[0.7ex]
\midrule
Marginal no common mechanism & Symmetric \(k\)-state model with independent edge probabilities at each site, the full cap \((k-1)/k\), and no clock. & Tuffley--Steel {[}11{]} give \(L^\star(T)=(M+S(T))\ln k\). Fixed-tree optimization is therefore exact by Fitch {[}12{]}. C.7 transfers parsimony approximation and hardness, distinguishing the normalized score from the full loss. A tighter probability cap or clock is a different model. \\[0.7ex]
\bottomrule
\end{tabular}
\endgroup
\end{landscape}

\begin{landscape}
\begingroup
\fontsize{9.5}{11.2}\selectfont
\setlength{\tabcolsep}{4pt}\setlength{\parskip}{0pt}
\renewcommand{\arraystretch}{1.05}
\textbf{Table C.1 (continued).} Ancestral likelihood.

\vspace{0.5em}
\begin{tabular}{@{}>{\raggedright\arraybackslash}p{(\linewidth - 4\tabcolsep) * \real{0.18}}>{\raggedright\arraybackslash}p{(\linewidth - 4\tabcolsep) * \real{0.35}}>{\raggedright\arraybackslash}p{(\linewidth - 4\tabcolsep) * \real{0.47}}@{}}
\toprule
\textbf{Problem and model} & \textbf{Optimization variables and objective} & \textbf{Algorithms, sources, and qualifications} \\ 
\midrule
Ancestral clipped binary CFN & Choose all ancestral sequences and shared edge probabilities in \([0,1/2]\). Removing the uniform root term and fitting probabilities gives \(J_2=\sum_eD_2\). & Max-product pruning replaces sums by maxima. Elias--Tuller {[}3, Theorem 4.8{]} give the fixed-tree PTAS and a topology factor two. C.2 spells out the clipped cost and endpoint arithmetic. The binary hardness construction of {[}21{]} has its optimum inside the clipped domain; C.7 derives a fixed gap. \\[0.7ex]
\midrule
Ancestral unclipped binary & Choose ancestral sequences and shared \(p_e\in[0,1]\). The fitted edge cost is \(M H_2(d_H/M)\), denoted \(D_{\rm u}\); its sum is \(J_{\rm u}\). & Alon et al.'s \(16/9\) topology guarantee concerns this objective {[}7{]}. Complementary sequences can have zero cost, unlike clipped CFN. C.2 applies the prescribed-tree metric method {[}2{]} after identifying zero-distance sequences. The formal binary model in {[}21{]} also uses this unclipped domain. \\[0.7ex]
\midrule
Ancestral symmetric \(k\)-state / JC69 & Fixed \(k\), uniform root, and shared \(p_e\in[0,(k-1)/k]\). Fitting probabilities gives \(J_k=\sum_eD_k\). & Symmetric max-product pruning uses the largest and second-largest child scores. Elias--Tuller {[}3, §5.2{]} extend their approximation algorithms to symmetric alphabets. C.3 improves the table's topology bound by a direct Steiner-tree consequence. C.7 extends the binary hardness calibration of {[}21{]}; {[}21{]} does not state that \(k\)-state theorem. \\[0.7ex]
\midrule
Ancestral reversible processes / phylogeography & Supplied irreducible reversible processes with their stationary root laws; choose ancestral sequences and branch durations. Use full joint loss \(J_{\rm full}\). Location can be another evolving coordinate. & The \(2\log_2 n\) bound of {[}8, Theorem 3{]} uses stationary-sequence and optimized-transition oracles and chooses the topology. Its edge score is a supremum over duration. C.6 gives the fixed-tree bound in that same oracle interpretation and explains finite output. C.7 obtains full-loss hardness from one fixed symmetric generator. \\[0.7ex]
\bottomrule
\end{tabular}
\endgroup
\end{landscape}

\begin{landscape}
\begingroup
\fontsize{9.5}{11.2}\selectfont
\setlength{\tabcolsep}{4pt}\setlength{\parskip}{0pt}
\renewcommand{\arraystretch}{1.05}
\textbf{Table C.1 (continued).} Parsimony, distance fitting, and Bayesian inference.

\vspace{0.5em}
\begin{tabular}{@{}>{\raggedright\arraybackslash}p{(\linewidth - 4\tabcolsep) * \real{0.18}}>{\raggedright\arraybackslash}p{(\linewidth - 4\tabcolsep) * \real{0.35}}>{\raggedright\arraybackslash}p{(\linewidth - 4\tabcolsep) * \real{0.47}}@{}}
\toprule
\textbf{Problem and model} & \textbf{Optimization variables and objective} & \textbf{Algorithms, sources, and qualifications} \\ 
\midrule
Unordered parsimony & Choose ancestral states, charging one for a change and zero otherwise. The same rule applies to every edge; optimize \(S\). & Fitch solves a supplied topology {[}12{]}. C.3 explains the component-based Steiner bound {[}7,18{]}. Binary instances already give the fixed gap {[}4{]}, so the unrestricted character model inherits it. \\[0.7ex]
\midrule
Parsimony with supplied costs & For the topology guarantee, each site's supplied state-cost matrix is a symmetric nonnegative metric, used on every edge. Optimize the sum \(S_c\). & Sankoff solves a fixed tree even for general dense state-cost matrices {[}14{]}. The topology bounds use the metric condition: sum the site metrics to obtain a sequence metric, then apply C.3. They are not claimed for arbitrary asymmetric or nonmetric costs. Unit binary costs supply hardness {[}4{]}. \\[0.7ex]
\midrule
Binary near-perfect parsimony & On \(d\) nonconstant binary coordinates, write the optimum as \(d+q\). Optimize the ordinary parsimony score. & Awasthi et al.~{[}13{]} return cost \(d+O(q^2)\) with probability at least \(1/2\). This is an additive guarantee for the near-perfect regime. General parsimony hardness is not asserted for a promise that \(q\) is small. Fixed-tree scoring remains Fitch's algorithm {[}12{]}. \\[0.7ex]
\midrule
Least-squares distance fitting & Supply a complete rational dissimilarity matrix and fit nonnegative edge lengths. Minimize \(\mathrm{SSE}=\sum_{u<v}(d_{uv}-d_T(u,v))^2\). & Fixed-topology fitting is a convex quadratic program {[}15{]}. The \(\ell_2\) tree-metric approximation of Ailon--Charikar {[}19{]} yields the stated squared-error ratio; C.8 gives the conversion and source scope. Neighbor joining {[}16{]} is a reconstruction method, not a guarantee for this optimization objective. \\[0.7ex]
\midrule
Bayesian inference & Supply a substitution model and priors. The targets are posterior distributions over ancestral states, parameters, or topologies, according to which quantities are observed. & With the tree and parameters fixed, tree message passing gives conditional ancestral marginals. Parameter and topology inference instead require integration or sampling guarantees, such as a specified error and runtime for MCMC {[}17{]}. An ML optimization ratio is not applicable. \\[0.7ex]
\bottomrule
\end{tabular}
\endgroup
\end{landscape}

\subsection{C.2. Symmetric likelihood and fixed-tree
approximation}\label{c.2.-symmetric-likelihood-and-fixed-tree-approximation}

Fix an alphabet of size \(k\ge2\), a uniform root distribution, and an
edge-specific probability shared across all sites of the explicit
alignment. Write \(c=(k-1)/k\). Fitting an edge between sequences
\(x,y\), with total change probability in \([0,c]\), gives \[
D_k(x,y)=M[H_2(u)+u\ln(k-1)],
\qquad u=\min\{d_H(x,y)/M,c\}.
\] This cost is a nondecreasing, concave function of Hamming distance
and vanishes at zero, so it is a metric. After fitting the
probabilities, a complete reconstruction has normalized ancestral loss
\(\sum_{uv\in E}D_k(x_u,x_v)\).

The unclipped binary model instead allows \(p_e\in[0,1]\) and has cost
\[
D_{\rm u}(x,y)=M H_2(d_H(x,y)/M).
\] This is a pseudometric {[}7{]}. Indeed, at a uniformly chosen site,
the mismatch indicator for \(x,y\) is determined by the mismatch
indicators for \(x,z\) and \(z,y\). Entropy subadditivity therefore
proves the triangle inequality. Its zero-distance pairs are equal or
complementary sequences. Identifying \(x\) with its complement gives a
metric. Leaf classes retain their observed representatives, and internal
representatives may be chosen freely without changing costs.

Both models permit exact optimization of a component with a fixed number
\(b_C\) of edges and prescribed boundary sequences. Some optimum uses
fitted probabilities from \[
\mathcal P_M=\{\min\{j/M,c\}:0\le j\le M\},
\] or \(\{j/M:0\le j\le M\}\) in the unclipped model. Enumerate these
vectors and, for each, maximize the complete likelihood at every site by
max-product pruning. There are at most \((M+1)^{b_C}\) vectors. Their
rational likelihood products can be compared exactly with polynomial bit
cost. The fixed-topology decomposition of {[}2{]} therefore applies to
both metrics, including the unclipped quotient. The clipped ancestral
PTAS and its symmetric extension are prior results of {[}3, Theorem 4.8
and §5.2{]}; this argument makes their probability cap and finite
implementation explicit. The unclipped fixed-tree PTAS follows from the
same decomposition.

For marginal likelihood, the comparison with ancestral likelihood is \[
F_k(p)\le J_k(p)\le2F_k(p).
\tag{C.1}
\] It holds for every feasible vector \(p\). To see this, apply Lemma 3
with \(f=-\ln(kP_p(z))\). Its entropy argument uses only binary
branching and the fact that each transition matrix is doubly stochastic,
so the alphabet size does not change the bound \(\mathcal E\le2f\). The
minimum transition loss of a compatible complete labeling lies between
\(f\) and \(\mathcal E\). Summing over sites proves (C.1). Thus the
prior ancestral PTAS already yields a \((2+\varepsilon)\) marginal
approximation through this comparison.

\textbf{Corollary 9 (Symmetric Small ML).} For every fixed \(k\ge2\),
Small ML in the symmetric \(k\)-state model admits a deterministic PTAS
for \(F_k\) and \(L\) on a prescribed rooted binary tree and explicit
alignment, with feasible rational output.

\textbf{Proof.} We give the changes required in Section 3. The
transition matrix is \[
K_p=\left(1-\frac{kp}{k-1}\right)I+
\frac{kp}{k-1}\frac{\mathbf1\mathbf1^\top}{k},
\] and composition obeys \[
K_pK_q=K_{p\oplus q},\qquad
p\oplus q=p+q-\frac{k}{k-1}pq\in[0,c].
\] Using \(D_k\) in the sequence-replacement argument therefore
preserves Lemmas 4--5 and their \(7/t\) bound. The root contribution is
now \(M\ln k\).

Let \(M_{\rm nc}\) count nonconstant sites. Such a site specifies
different states at two leaves. Their joint probability is at most
\(1/k^2\), because the path transition is another capped symmetric
channel. Hence \(F_k^\star\ge M_{\rm nc}\ln k\). If every site is
constant, \(p_e=0\) is exact. Since every transition probability is at
most \(1-p_e\), the partial-loss bound \(M\sum_e p_e\le2G_S\) also
remains valid.

For \(0<\varepsilon\le1\), keep \(t=\lceil14/\varepsilon\rceil\), but
set \[
\tau=\frac{\varepsilon}{32k},\qquad
\lambda=\frac{\varepsilon M_{\rm nc}}{32kMb},\qquad
\mathcal G=\{0,c\}\cup
\{\lambda(1+\tau)^j:j\ge0,\ \lambda(1+\tau)^j<c\}.
\] The derivative of \(-\ln(1-p)\) is at most \(k\) on \([0,c]\). Upward
rounding thus gives \[
G_S(p')\le(1+2k\tau)G_S(p)+kMb\lambda
=(1+\varepsilon/16)G_S(p)+\varepsilon M_{\rm nc}/32.
\] Combining this with the \((1+\varepsilon/2)\) partial witness and
\(\ln k\ge\ln2>1/2\) gives \[
F_k(\widehat p)\le
\left(1+\frac{21\varepsilon}{32}\right)F_k^\star.
\] The dynamic program sums \(k\) root messages in the top component.
The grid contains \[
O\!\left(\frac{k}{\varepsilon}\log\frac{kMb}{\varepsilon M_{\rm nc}}\right)
\] entries, and component height is unchanged. All transition factors
are rational, so Appendix A's product arithmetic remains polynomial for
fixed \(k,\varepsilon\). Adding \(M\ln k\) proves the claim for \(L\).
\(\square\)

This corollary extends the approximation argument; it does not extend
Theorem 2's binary hardness reduction.

\subsection{C.3. Consequences of Steiner
approximation}\label{c.3.-consequences-of-steiner-approximation}

The published ancestral topology guarantees include factor two for the
clipped symmetric model {[}3{]} and \(16/9\) for the unclipped binary
objective {[}7{]}. Combining exact component optimization with the later
Steiner approximation {[}18{]} gives the following stronger consequence.

\textbf{Corollary 10 (Topology approximation).} For fixed \(k\) and
\(\varepsilon>0\), the clipped symmetric ancestral objective has a
deterministic \((\ln4+\varepsilon)\)-approximation for topology search.
The same holds for the unclipped binary ancestral objective. Symmetric
marginal likelihood has a deterministic
\((2\ln4+\varepsilon)\)-approximation for both \(F_k\) and \(L\).

\textbf{Proof.} Let \(\operatorname{OPT}\) denote the minimum normalized
ancestral loss over trees on the observed sequences. Fix \(r\ge2\). For
every subset of at most \(r\) observed sequences, compute an optimal
full component: a tree whose prescribed leaves are that subset.
Enumerate its binary topologies and use the exact solver in C.2. This
takes polynomial time for fixed \(r,k\). Form a graph \(H\) from the
union of these components, sharing their prescribed terminal vertices.
Internal vertices may remain separate copies when their sequences agree.
Thus \(H\) has polynomial size, although the space of possible ancestral
sequences has size \(k^M\).

The restricted-Steiner theorem quoted in {[}18, Theorem 1{]} gives a
tree assembled from components with at most \(r\) terminals, of total
cost at most \(\rho_r\operatorname{OPT}\), where \[
\rho_r\le1+\frac1{\lfloor\log_2r\rfloor}.
\] Replacing each component by the stored optimum for its terminal set
gives a connected subgraph of \(H\). Consequently
\(\operatorname{OPT}(H)\le\rho_r\operatorname{OPT}\). This construction
avoids enumerating the full sequence graph.

For explicit alignments, every positive entropy edge cost is at least
\(\ln2\); zero costs are recognizable exactly from the disagreement
count. Approximate each positive cost \(d_e\) by a rational upper bound
\[
d_e\le\widehat d_e\le d_e+\delta/2\le(1+\delta)d_e,
\] and preserve zeros. Range-reduced logarithm series compute these
bounds with polynomial precision. Contract zero-cost edges before
applying {[}18{]} and restore them afterward. The rounded costs need not
remain a metric: the deterministic algorithm of {[}18{]} applies to
weighted graphs. Its output has true cost at most \[
(\ln4+\delta)(1+\delta)\rho_r\operatorname{OPT}.
\] For \(0<\varepsilon\le1\), choose \(\delta=\varepsilon/16\) and fixed
\(r\) with \(\rho_r\le1+\delta\). The resulting factor is less than
\(\ln4+\varepsilon\).

Remove unnecessary nonterminal leaves. Convert the remaining tree to a
binary phylogeny by metric shortcutting at degree-two nonterminals and
by adding zero-cost copies to resolve branching and retain named leaves.
In the unclipped model, complementary copies have zero cost and use
\(p_e=1\); identical copies use \(p_e=0\). Insert a root by duplicating
an edge endpoint. Fitting the remaining edges returns rational feasible
parameters. Finally, (C.1), applied with ancestral accuracy
\(\varepsilon/2\), gives the marginal factor \(2\ln4+\varepsilon\).
Adding the uniform-root term preserves all factors for the full losses.
\(\square\)

The same construction gives \(\ln4+\varepsilon\) for parsimony with
rational symmetric metric state costs, provided each site's cost matrix
is used on every edge. The sum of site costs is a metric on sequences,
and each fixed-size component is solved by Sankoff dynamic programming
{[}14{]}. Rational input costs require no entropy approximation. These
are consequences of the cited algorithms, rather than claims made in
their original phylogenetic statements.

\subsection{C.4. Supplied generators, clocks, and finite rate
mixtures}\label{c.4.-supplied-generators-clocks-and-finite-rate-mixtures}

The generator row supplies a rational irreducible reversible generator
\(Q\) on \(k\ge2\) states, with stationary root distribution \(\pi\),
and fits nonnegative finite branch durations \(t_e\), without an upper
bound. The generator is shared across sites. Jointly fitting a generator
is a different problem. Write \(P_{T,t}(z)\) for a leaf-pattern
probability. We compare the full marginal loss \(L\) with its infimum
\(L^\star\), which need not be attained at finite durations.

\textbf{Proposition 11 (Supplied reversible generators).} For every
rational \(0<\varepsilon\le1\), this problem has an
\((n+\varepsilon)\)-approximation returning rational durations, on
either a prescribed or an inferred topology.

\textbf{Proof.} Stationarity gives \(P_{T,t}(z^{(i)})\le\pi(z_v^{(i)})\)
for every leaf \(v\). Hence, with \(\pi_{\min}=\min_a\pi_a\) and
\(\pi_{\max}=\max_a\pi_a\), \[
C:=\sum_{i,v}-\ln\pi(z_v^{(i)})\le nL^\star,
\qquad L^\star\ge M(1-\pi_{\max}).
\] Let \(q_{\min}\) be the smallest positive off-diagonal entry of
\(Q\), and set \(\lambda_0=\pi_{\min}q_{\min}/k\). This bounds the
spectral gap of \(-Q\) from below. Indeed, paths in a spanning tree of
the support of \(Q\) give \[
\operatorname{Var}_\pi(f)
\le\frac{k-1}{2\pi_{\min}q_{\min}}\,
\mathcal E_Q(f,f),
\qquad
\mathcal E_Q(f,f)=\frac12\sum_{a,b}\pi_aQ_{ab}(f_a-f_b)^2.
\] The variational characterization of the gap and reversible spectral
decomposition therefore imply \[
\left|\frac{(e^{tQ})_{ab}}{\pi_b}-1\right|
\le\frac{e^{-\lambda_0t}}{\pi_{\min}}.
\] Choose \[
\eta=\frac{\varepsilon(1-\pi_{\max})}{2n},
\qquad T_0=\frac1{\lambda_0\eta\pi_{\min}}.
\] Since \(e^{-x}\le1/x\), every entry of \(e^{T_0Q}\) is at least
\((1-\eta)\pi_b\). Give every pendant edge duration \(T_0\) and every
internal edge duration zero. Summing over ancestors yields \[
P_{\rm out}(z^{(i)})\ge(1-\eta)^n\prod_v\pi(z_v^{(i)}).
\] Using \(-\ln(1-\eta)\le2\eta\), we obtain \[
L_{\rm out}\le C+2nM\eta\le(n+\varepsilon)L^\star.
\] The stationary distribution is computable by rational linear algebra.
All constructed durations have polynomial encoding length, and their
computation is polynomial in the input length and
\(\log(1/\varepsilon)\). The construction preserves any supplied
topology. \(\square\)

For the strict CFN clock, all leaves have age zero; internal ages
respect ancestry, with ties allowed and no upper age bound. Assigning
age \(T_0\) to every internal vertex implements the construction. Shared
node ages couple the branch durations, so the independent-edge PTAS does
not directly impose this clock. Chor--Tuller {[}10, Theorem 3.6{]} prove
exact topology-search hardness for this contemporaneous two-state clock.

For a finite rate mixture, supply positive rational rates \(r_c\) and
weights \(w_c\), with \(\sum_cw_c=1\). One latent category applies to
every edge at a site: \[
P_{T,t}(z)=\sum_{c=1}^s w_cP_{T,r_ct}(z).
\] Pendant duration \(T_0/\min_cr_c\) gives the same bound within every
category and therefore after averaging. This formulation supplies the
rates and weights and excludes a zero-rate category. Fitting them
jointly defines another problem. Topology search is NP-hard for the
class allowing \(s=1\), already with supplied JC69 {[}9{]}; the
supplied-generator class has the same restriction.

For supplied-generator classes permitting \(k=2\), and mixture classes
permitting \(k=2,s=1\), a feasible rational-duration FPTAS for \(L\)
would imply \(RP=NP\). Use \[
Q=\begin{pmatrix}-1&1\\1&-1\end{pmatrix},
\qquad p_e(t_e)=\frac{1-e^{-2t_e}}2.
\] Finite rational durations approximate every point of the closed CFN
probability box, so their infimum equals its minimum. Run the assumed
FPTAS at accuracy \(\varepsilon/2\), then round each probability upward
to a rational \(p_e\le\widehat p_e\le\min\{p_e+\delta,1/2\}\), where
\(\delta=\varepsilon/(16b)\). Flip factors increase, while a nonflip
loses at most \[
\ln\frac{1-p_e}{1-\widehat p_e}\le2\delta.
\] This comparison survives summation over ancestors. Since
\(L^\star\ge M\ln2\), \[
L(\widehat p)\le(1+\varepsilon/2)L^\star+2Mb\delta
\le\left(1+\frac\varepsilon2+\frac\varepsilon{8\ln2}\right)L^\star
<(1+\varepsilon)L^\star.
\] Certified rounding takes polynomial time: sufficiently large
durations permit \(\widehat p_e=1/2\), and the remaining exponentials
require polynomial precision on an interval of length
\(O(\log(1/\delta))\). Theorem 2 applies. This restriction argument
establishes neither fixed-four-state GTR hardness nor clock hardness for
parameter fitting.

\subsection{C.5. Known CFN partitions}\label{c.5.-known-cfn-partitions}

Let partition \(j\) contain \(M_j\) observed sites. Each partition has
an independent edge-probability vector, while all partitions use one
topology. The losses add: \(F=\sum_jF_j\) and \(L=\sum_jL_j\); write
\(J=\sum_jJ_{2,j}\) for their combined normalized ancestral loss.
Applying Theorem 1 independently gives a fixed-tree PTAS. Theorem 2's
FPTAS obstruction follows by allowing one partition.

For topology search, use the metric \[
D_{\rm part}(x,y)=\sum_j M_jH_2\!\left(
\min\{d_H(x^{[j]},y^{[j]})/M_j,1/2\}\right),
\] where \(x^{[j]}\) is the restriction of sequence \(x\) to partition
\(j\). It is a sum of clipped entropy metrics. Applying the posterior
reconstruction bound within each partition of a common optimal tree
gives \(J^\star\le2F^\star\).

The constant-terminal component algorithm of C.3 applies to this sum
metric. For each terminal subset, enumerate its constant-size component
topologies. On each topology, optimize the partitions independently and
sum their costs; then choose the topology minimizing that sum. Their
ancestral coordinates assemble into full sequences on the same internal
vertices. This computes an exact component in polynomial time for each
fixed terminal bound. The rational cost approximation in C.3 also
applies because every positive sum-metric cost is at least \(\ln2\).

The resulting ancestral factor \(\ln4+\varepsilon/2\) gives marginal
factor \(2\ln4+\varepsilon\) for both \(F\) and \(L\). All partitions
share the returned topology. Roch's normalized fixed gap and exact
topology-search hardness {[}4{]} transfer through the one-partition
restriction.

\subsection{C.6. Reversible ancestral likelihood and
phylogeography}\label{c.6.-reversible-ancestral-likelihood-and-phylogeography}

Let \(P_i^t\) and \(\pi_i\) be the supplied finite-state reversible
transition kernel and stationary distribution at coordinate \(i\),
including a location coordinate when present, as in {[}8{]}. For full
sequences, define \[
\Phi(x)=-\ln\prod_i\pi_i(x_i),
\qquad c(x,y)=-\ln\sup_t\prod_iP_i^t(x_i,y_i).
\] The full ancestral objective includes \(\Phi\) at the root and \(c\)
on each edge. The oracles in {[}8{]} evaluate these costs; its topology
algorithm gives factor \(2\log_2n\).

A factor \(n-1\) holds on any prescribed topology. Choose an observed
sequence \(r\), assign it to every internal vertex, use duration zero
internally and on the pendant edge of \(r\), and optimize the other
pendant transitions. To compare with any complete reconstruction, reroot
it at leaf \(r\); reversibility preserves its joint probability. Along
the path from \(r\) to another observed sequence \(y\), the semigroup
identity gives \[
P_i^{\sum_et_e}(r_i,y_i)
\ge\prod_{e\text{ on the path}}P_i^{t_e}(x_{e^-,i},x_{e^+,i}).
\] Here \(e^-,e^+\) are the endpoints ordered along the path.
Multiplication over sites uses the same summed duration. Thus \(c(r,y)\)
is at most the path's transition loss. Summing over the other \(n-1\)
leaves charges each edge at most \(n-1\) times and the nonnegative root
loss once. Taking the infimum proves the bound.

This guarantee uses exact optimized-edge costs, as in {[}8{]}. A
supremum attained only at infinite time is interpreted in the model's
closure. Finite feasible durations require attaining-duration oracles or
certified approximate durations, with corresponding approximation slack.

\subsection{C.7. Topology hardness and the choice of
loss}\label{c.7.-topology-hardness-and-the-choice-of-loss}

All hardness statements here concern joint optimization of the topology
and its edge parameters. Addario-Berry et al.~{[}21{]} prove exact
NP-hardness of binary ancestral ML. We derive the following
approximation gap from their canonical-cost lemma and bounded-degree
Vertex Cover hardness {[}22{]}.

\textbf{Corollary 12 (Ancestral topology hardness).} For every fixed
\(k\ge2\), some constant \(c>0\) makes \((1+c)\)-approximation NP-hard
for both \(J_k\) and the full ancestral loss
\(J_{\rm full}=J_k+M\ln k\), with a uniform root and shared symmetric
edge probabilities in \([0,(k-1)/k]\). The binary conclusion also holds
for the unclipped domain \([0,1]\). Full-loss hardness already holds for
one fixed supplied JC69 generator.

\textbf{Proof.} Let a graph have \(v\) vertices, \(m\) edges, and
minimum vertex-cover size \(\tau\). Replace each vertex by two
nonadjacent twins and each edge by all four edges between its twin
pairs. The expanded graph has \(2v\) vertices, \(4m\) edges, and cover
number \(2\tau\). Following {[}21{]}, form an alignment of length
\(M=2v\), with one all-zero taxon and, for each expanded edge, one taxon
having ones at its endpoints and zeros elsewhere. Define \[
a=4m+2\tau,\qquad C_M=M H_2(1/M).
\] Here \(H_2\) is binary entropy with natural logarithms. For
sufficiently large \(v\), Lemmas 3--4 of {[}21{]} give minimum unclipped
binary entropy cost \(aC_M\), attained by a canonical reconstruction
with exactly one disagreement on each positive-cost edge. Its fitted
probabilities are \(0\) or \(1/M\), so the clipped binary optimum is
identical. Duplicate sequences on zero-cost edges attach named leaves
and resolve high-degree vertices into a binary tree.

For the clipped symmetric \(k\)-state model, mapping every ancestral
symbol other than 0 and 1 to 0 cannot decrease likelihood: a changed
transition either remains a change or becomes a more probable match.
Thus an optimum uses binary ancestral sequences. Let \(D_k(d)\) be the
fitted cost of an edge with \(d\) disagreements. Relaxing its
probability cap gives \[
D_k(d)\ge M H_2(d/M)+d\ln(k-1).
\] Entropy concavity gives \(M H_2(d/M)\le dC_M\) for integer \(d\ge1\).
Hence every binary reconstruction, with edge disagreement counts
\(d_e\), satisfies \[
\sum_e M H_2(d_e/M)\ge aC_M,
\qquad \sum_e d_e\ge a.
\] Both lower bounds are attained by the same canonical reconstruction.
Writing \(C_{M,k}=C_M+\ln(k-1)\), we obtain \[
J_k^\star=(4m+2\tau)C_{M,k},\qquad
J_{\rm full}^\star=2v\ln k+(4m+2\tau)C_{M,k}.
\tag{C.2}
\]

Use a bounded-degree Vertex Cover gap {[}22{]}: for fixed constants
\(0<\alpha<\beta<1\) and \(\Delta\), distinguishing \(\tau\le\alpha v\)
from \(\tau\ge\beta v\) is NP-hard on graphs of maximum degree
\(\Delta\). The YES upper threshold \(A=2v\ln k+(4m+2\alpha v)C_{M,k}\)
and NO lower threshold \(B=A+2(\beta-\alpha)vC_{M,k}\) satisfy \[
\frac BA\ge1+\frac{\beta-\alpha}{1+\Delta+\alpha},
\] because \(m\le\Delta v/2\) and \(C_{M,k}\ge\ln k\). Removing the root
term only increases this separation. Choose \(1+c\) strictly below this
ratio. The alignment has polynomial size, and the fixed separation
permits polynomial-precision evaluation of the logarithmic thresholds.
If an algorithm returns only fitted parameters and a topology,
max-product pruning recovers an ancestral assignment with the same
objective value.

Finally, fix \(Q_{ii}=-1\) and \(Q_{ij}=1/(k-1)\) for \(i\ne j\). This
generator is irreducible and reversible with uniform stationary
distribution, and its total change probability at duration \(t\) is \[
p(t)=\frac{k-1}{k}\left(1-e^{-kt/(k-1)}\right).
\] Every canonical probability \(1/M\) has a finite realizing duration;
zero-cost copies use \(t=0\). Thus (C.2) also holds for this fixed
generator, which is JC69 when \(k=4\). This establishes hardness for the
supplied-generator class. \(\square\)

For \textbf{no common mechanism}, Tuffley--Steel {[}11{]} give
\(F_k^\star(T)=S(T)\ln k\) and \(L^\star(T)=(M+S(T))\ln k\). A parsimony
approximation therefore gives the same factor for either loss. For
hardness, remove constant sites, so \(M\le S^\star\). A
\((1+\varepsilon)\)-approximation to full loss, followed by exact
fitting on its returned topology \(T\), implies \[
S(T)\le S^\star+\varepsilon(M+S^\star)
\le(1+2\varepsilon)S^\star.
\] The binary parsimony gap {[}4{]} consequently gives a fixed gap for
both losses. It applies to every fixed \(k\ge2\), since mapping unused
ancestral symbols to 0 preserves binary observations and cannot increase
parsimony cost.

For \textbf{shared-parameter marginal ML}, Roch's {[}4{]} gap concerns
normalized \(F\). Its root subtraction cannot automatically be reversed
for approximation ratios. Corollary 12 concerns ancestral loss; the
comparison \(F_k\le J_k\le2F_k\) does not transfer its fixed gap to
marginal loss.

\subsection{C.8. Distance fitting and Bayesian
inference}\label{c.8.-distance-fitting-and-bayesian-inference}

For rational observed dissimilarities \(d_{uv}\), the distance-fitting
objective is \[
\operatorname{SSE}(T,\ell)=\sum_{u<v}
\left(d_{uv}-\sum_{e\in\operatorname{path}_T(u,v)}\ell_e\right)^2,
\qquad\ell_e\ge0.
\] On a fixed topology this is a rational convex quadratic program and
is solvable exactly in polynomial time {[}15{]}. For topology search,
Ailon--Charikar {[}19{]} give an
\(O((\log n\log\log n)^{1/p})\)-approximation in the \(\ell_p\) norm.
Taking \(p=2\) and squaring the guarantee gives factor
\(O(\log n\log\log n)\) for SSE. Zero-length edges permit named leaves
and binary refinement without changing distances. Neighbor joining
{[}16{]} is a reconstruction procedure; this guarantee comes from
tree-metric fitting.

Exact \(\ell_2\) topology fitting is stated to be NP-hard in {[}19,
§1.1{]}. Day's exact-hardness theorem {[}20{]} uses positive
half-integer edge lengths and concerns that discrete formulation.

The Bayesian entries are marked ``not applicable'' because their target
is a posterior distribution. Conditional ancestral marginals can be
computed by tree message passing at supplied parameters. Integration
over parameters and sampling over parameters or topologies require a
specified error measure and runtime guarantee {[}17{]}, rather than a
ratio for the objectives above.

\section{References}\label{references}
\begingroup\small\raggedright\setlength{\parskip}{0.45em}
\begin{samepage}
{[}1{]} Joseph Felsenstein. Evolutionary trees from DNA sequences: A
maximum likelihood approach. \emph{Journal of Molecular Evolution}
17:368--376, 1981.
\href{https://doi.org/10.1007/BF01734359}{doi:10.1007/BF01734359}.
\par\end{samepage}

\begin{samepage}
{[}2{]} Lusheng Wang, Tao Jiang, and Eugene L. Lawler. Approximation
algorithms for tree alignment with a given phylogeny.
\emph{Algorithmica} 16:302--315, 1996.
\href{https://www.cs.cityu.edu.hk/~lwang/research/alg96.pdf}{Author
manuscript}.
\par\end{samepage}

\begin{samepage}
{[}3{]} Isaac Elias and Tamir Tuller. Reconstruction of ancestral
genomic sequences using likelihood. \emph{Journal of Computational
Biology} 14(2):216--237, 2007.
\href{https://doi.org/10.1089/cmb.2006.0101}{doi:10.1089/cmb.2006.0101}.
\href{https://www.cs.tau.ac.il/~tamirtul/Pub/AML.pdf}{Author
manuscript}.
\par\end{samepage}

\begin{samepage}
{[}4{]} Sébastien Roch. A short proof that phylogenetic tree
reconstruction by maximum likelihood is hard. \emph{IEEE/ACM
Transactions on Computational Biology and Bioinformatics} 3(1):92--94,
2006. \href{https://arxiv.org/abs/math/0504378}{arXiv:math/0504378}.
\par\end{samepage}

\begin{samepage}
{[}5{]} David Clancy Jr., Hanbaek Lyu, and Sebastien Roch. Likelihood
landscape of binary latent model on a tree. arXiv:2501.17622, 2025;
revised May 2026. \href{https://arxiv.org/abs/2501.17622}{Preprint}.
\par\end{samepage}

\begin{samepage}
{[}6{]} David Clancy Jr., Hanbaek Lyu, and Sebastien Roch. Sample
complexity of branch-length estimation by maximum likelihood. In
\emph{Proceedings of the 42nd International Conference on Machine
Learning}, PMLR 267:11042--11063, 2025.
\href{https://proceedings.mlr.press/v267/clancy25a.html}{Proceedings}.
\par\end{samepage}

\begin{samepage}
{[}7{]} Noga Alon, Benny Chor, Fabio Pardi, and Anat Rapoport.
Approximate maximum parsimony and ancestral maximum likelihood.
\emph{IEEE/ACM Transactions on Computational Biology and Bioinformatics}
7(1):183--187, 2010.
\href{https://doi.org/10.1109/TCBB.2008.13}{doi:10.1109/TCBB.2008.13}.
\href{https://web.math.princeton.edu/~nalon/PDFS/acpr18.pdf}{Author
manuscript}.
\par\end{samepage}

\begin{samepage}
{[}8{]} Mohammad-Hadi Foroughmand-Araabi, Sama Goliaei, and Kasra
Alishahi. An Approximation Algorithm for Ancestral Maximum-Likelihood
and Phylogeography Inference Problems under Time Reversible Markov
Evolutionary Models. arXiv:2308.06561, 2023.
\href{https://arxiv.org/abs/2308.06561}{Preprint}.
\par\end{samepage}

\begin{samepage}
{[}9{]} Benny Chor and Tamir Tuller. Maximum likelihood of evolutionary
trees is hard. RECOMB, 2005.
\href{https://www.cs.tau.ac.il/~tamirtul/Pub/ML.pdf}{Expanded author
manuscript}.
\par\end{samepage}

\begin{samepage}
{[}10{]} Benny Chor and Tamir Tuller. Maximum likelihood of evolutionary
trees: hardness and approximation. \emph{Bioinformatics} 21(suppl.
1):i97--i106, 2005.
\href{https://doi.org/10.1093/bioinformatics/bti1027}{DOI}.
\par\end{samepage}

\begin{samepage}
{[}11{]} Chris Tuffley and Mike Steel. Links between maximum likelihood
and maximum parsimony under a simple model of site substitution.
\emph{Bulletin of Mathematical Biology} 59(3):581--607, 1997.
\href{https://doi.org/10.1007/BF02459467}{DOI}.
\par\end{samepage}

\begin{samepage}
{[}12{]} Walter M. Fitch. Toward defining the course of evolution:
Minimum change for a specific tree topology. \emph{Systematic Zoology}
20(4):406--416, 1971.
\href{https://doi.org/10.1093/sysbio/20.4.406}{DOI}.
\par\end{samepage}

\begin{samepage}
{[}13{]} Pranjal Awasthi, Avrim Blum, Jamie Morgenstern, and Or Sheffet.
Additive Approximation for Near-Perfect Phylogeny Construction.
arXiv:1206.3334, 2012. See §1.1 for the application of the
Byrka--Grandoni--Rothvoß--Sanità Steiner approximation.
\href{https://arxiv.org/abs/1206.3334}{Preprint}.
\par\end{samepage}

\begin{samepage}
{[}14{]} David Sankoff. Minimal mutation trees of sequences. \emph{SIAM
Journal on Applied Mathematics} 28(1):35--42, 1975.
\href{https://doi.org/10.1137/0128004}{DOI}.
\par\end{samepage}

\begin{samepage}
{[}15{]} M. K. Kozlov, S. P. Tarasov, and L. G. Khachiyan. The
polynomial solvability of convex quadratic programming. \emph{USSR
Computational Mathematics and Mathematical Physics} 20(5):223--228,
1980. \href{https://doi.org/10.1016/0041-5553(80)90098-1}{DOI}.
\par\end{samepage}

\begin{samepage}
{[}16{]} Naruya Saitou and Masatoshi Nei. The neighbor-joining method: A
new method for reconstructing phylogenetic trees. \emph{Molecular
Biology and Evolution} 4(4):406--425, 1987.
\href{https://doi.org/10.1093/oxfordjournals.molbev.a040454}{DOI}.
\par\end{samepage}

\begin{samepage}
{[}17{]} John P. Huelsenbeck and Fredrik Ronquist. MRBAYES: Bayesian
inference of phylogenetic trees. \emph{Bioinformatics} 17(8):754--755,
2001. \href{https://doi.org/10.1093/bioinformatics/17.8.754}{DOI}.
\par\end{samepage}

\begin{samepage}
{[}18{]} Jarosław Byrka, Fabrizio Grandoni, Thomas Rothvoß, and Laura
Sanità. Steiner tree approximation via iterative randomized rounding.
\emph{Journal of the ACM} 60(1), 2013.
\href{https://doi.org/10.1145/2432622.2432628}{doi:10.1145/2432622.2432628}.
\href{https://people.idsia.ch/~grandoni/Pubblicazioni/BGRS12jacm.pdf}{Author
manuscript}.
\par\end{samepage}

\begin{samepage}
{[}19{]} Nir Ailon and Moses Charikar. Fitting tree metrics:
Hierarchical clustering and phylogeny. \emph{SIAM Journal on Computing}
40(5):1275--1291, 2011.
\href{https://doi.org/10.1137/100806886}{doi:10.1137/100806886}.
\href{https://web.eecs.umich.edu/~pettie/theoryseminar/papers/Ailoncharikar-fitting-tree-metrics.pdf}{Conference
manuscript}.
\par\end{samepage}

\begin{samepage}
{[}20{]} William H. E. Day. Computational complexity of inferring
phylogenies from dissimilarity matrices. \emph{Bulletin of Mathematical
Biology} 49(4):461--467, 1987.
\href{https://doi.org/10.1007/BF02458863}{doi:10.1007/BF02458863}.
\par\end{samepage}

\begin{samepage}
{[}21{]} Louigi Addario-Berry, Benny Chor, Mike Hallett, Jens Lagergren,
Alessandro Panconesi, and Todd Wareham. Ancestral maximum likelihood of
evolutionary trees is hard. \emph{Journal of Bioinformatics and
Computational Biology} 2(2):257--271, 2004.
\href{https://doi.org/10.1142/S0219720004000557}{doi:10.1142/S0219720004000557}.
\href{https://www.cs.tau.ac.il/research/benny.chor/AML.pdf}{Author
manuscript}.
\par\end{samepage}

\begin{samepage}
{[}22{]} Piotr Berman and Marek Karpinski. On some tighter
inapproximability results. In \emph{ICALP 1999}, Lecture Notes in
Computer Science 1644:200--209, 1999.
\href{https://doi.org/10.1007/3-540-48523-6_17}{doi:10.1007/3-540-48523-6\_17}.
\href{https://archive.dimacs.rutgers.edu/TechnicalReports/abstracts/1999/99-23.html}{Technical
report}.
\par\end{samepage}
\endgroup

\end{document}